\documentclass[twocolumn,aps,prl] {revtex4} 
\usepackage{graphicx}
\begin{document}
\pagestyle{empty} 
\title{Rubber friction: role of the flash temperature}
\author{B.N.J. Persson}
\affiliation{IFF, FZ-J\"ulich, 52425 J\"ulich, Germany}

\begin{abstract}
When a rubber block is sliding on a hard rough substrate, the substrate
asperities will exert time-dependent deformations of the rubber surface
resulting in viscoelastic energy dissipation in the rubber, which gives a contribution
to the sliding friction. Most surfaces of solids have roughness on many different length scales,
and when calculating the friction force it is necessary to include the viscoelastic deformations
on all length scales. The energy dissipation will result in local heating of the rubber.
Since the viscoelastic properties of rubber-like materials 
are extremely strongly temperature dependent, it is
necessary to include the local temperature increase in the analysis. At very low sliding velocity
the temperature increase is negligible because of heat diffusion, but already for velocities of order
$10^{-2} \ {\rm m/s}$ the local heating may be very important. Here I study the influence of the local
heating on the rubber friction, and I show that in a typical case the temperature increase
results in a decrease in rubber friction  
with increasing sliding velocity for $v > 0.01 \ {\rm m/s}$. This may result in stick-slip
instabilities, and is of crucial importance in many practical applications, e.g., for the tire-road
friction, and in particular for ABS-breaking systems.  
\end{abstract}
\maketitle


\vskip 0.5cm
{\bf 1 Introduction}

Rubber friction is of extreme practical importance, e.g., in the context 
of tires, wiper blades, conveyor belts and seals\cite{Moore}.
Rubber friction on smooth substrates, e.g., on smooth glass 
surfaces, has two contributions, namely an adhesive (surface) and a hysteretic 
(bulk) contribution\cite{Moore,PerssonSS}. The adhesive contribution
results from the attractive binding forces between the rubber 
surface and the substrate. Surface forces are often dominated by 
weak attractive van der Waals interactions.
For very smooth substrates, because of the low elastic moduli of rubber-like materials, 
even when the applied squeezing force is very gentle this weak attraction 
may result in a nearly complete contact at the 
interface\cite{Persson,P1,Sophie}, 
leading to the large sliding friction force usually observed\cite{Grosch}. 
For rough surfaces, on the other hand, the adhesive contribution to rubber 
friction will be much smaller because of the small contact area. The actual 
contact area between a tire and the road surface, for example, is 
typically only $\sim 1 \%$ of the 
nominal footprint contact area\cite{P8,PV,Heinrich}. 
Under these conditions the bulk (hysteretic) friction mechanism 
is believed to prevail 
\cite{P8,Heinrich}.
For example, the exquisite sensitivity of tire-road friction to temperature 
just reflects the strong temperature dependence of the viscoelastic bulk 
properties of rubber.

When a rubber block is slid on a hard rough substrate the surface asperities of the
substrate will exert fluctuating forces on the rubber surface which, because of the
internal friction of the rubber, will result in 
energy transfer from the translational motion of the block
into the irregular thermal motion. This will result in a  
contribution to the friction force
acting on the rubber block.
The energy dissipation will result in local heating of the rubber.
Since the viscoelastic properties of rubber-like materials 
are extremely strongly temperature dependent, it is
necessary to include the local temperature increase when calculating the friction force. 
At very low sliding velocity
the temperature increase is negligible because of heat diffusion, but already for velocities of order
$10^{-2} \ {\rm m/s}$ the local heating may become very important. Here I study the influence of the local
heating on the rubber friction.
I show that in a typical case the temperature increase will result in a friction 
which decreases with increasing sliding velocity for $v > 0.01 \ {\rm m/s}$. 
This may result in stick-slip
instabilities, and is of crucial importance in many practical applications, e.g., for the tire-road
friction, and in particular for ABS-breaking systems.  

\vskip 0.5cm
{\bf 2 Surface roughness and macro-asperity contact}

The influence of surface roughness on
rubber friction
is mainly determined by the surface roughness
power spectrum $C(q)$ defined by
$$C(q)= {1\over (2\pi )^2 } \int d^2x \ \langle h({\bf x})h({\bf 0})\rangle 
e^{-i{\bf q}\cdot {\bf x}} 
\eqno(1)$$
Here $h({\bf x})$ is the substrate height measured from the average plane defined so that
$\langle h \rangle = 0$. The $\langle \ldots \rangle $ stand for ensemble averaging,
or averaging over the surface area. We have assumed that the statistical properties
of the substrate are translational invariant and isotropic so that $C(q)$ only depend on the
magnitude $q=|{\bf q}|$ of the wave vector ${\bf q}$. Note that from (1) follows    
$$ \langle h({\bf x})h({\bf 0})\rangle = \int d^2q \ C(q) e^{i{\bf q}\cdot {\bf x}} $$
so that the root-mean-square (rms) roughness amplitude 
$h_0 = \langle h^2\rangle^{1/2}$ is determined by 
$$ \langle h^2 \rangle = \int d^2q \ C(q) = 2\pi \int_0^\infty dq \ q C(q)\eqno(2)  $$
In reality, there will always be an upper, $q_1$, and a lower, $q_L$, limit to the $q$-integral in (2). Thus,
the largest possible wave vector will be of order $2\pi /a$, where $a$ is some lattice constant,
and the smallest possible wave vector is of order $2\pi /L$ where $L$ is the linear size of the surface.
 
\begin{figure}
\includegraphics[width=0.35\textwidth]{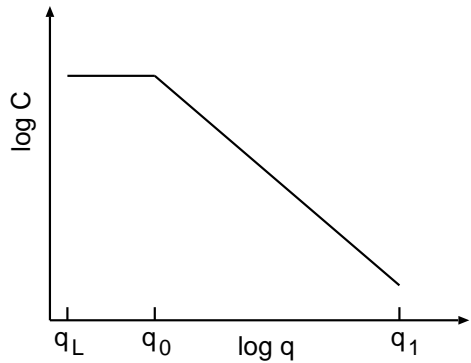}
\caption{\label{Cq1} 
Surface roughness power spectrum of a surface which is self affine fractal for
$q_1>q>q_0$. The long-distance roll-off wave vector $q_0$ and the short distance cut-off
wave vector $q_1$ depend on the system under consideration. The slope of 
the ${\rm log} C-{\rm log}q$ relation for $q > q_0$ determines the fractal 
exponent of the surface. The lateral size $L$ of the surface (or of the studied surface region)
determines the smallest possible wave vector $q_L=2\pi /L$.} 
\end{figure}

Many surfaces tend to be nearly self-affine fractal. A self-affine fractal
surface has the property that if part of the surface is magnified, with a magnification
which in general is appropriately different in the perpendicular direction to the surface as compared
to the lateral directions, then the surface ``looks the same'', i.e., the statistical
properties of the surface are invariant under the scale transformation. 
For a self-affine
surface the power spectrum has the power-law behavior 
$$C(q) \sim q^{-2(H+1)},$$
where the Hurst exponent $H$ is related to the fractal dimension $D_{\rm f}$ of the surface via
$H=3-D_{\rm f}$. Of course, for real surfaces this relation only holds in some finite 
wave vector region $q_0 < q < q_1$, and in a typical case $C(q)$ has the form shown
in Fig.~\ref{Cq1}. Note that in many cases there is a roll-off wavevector $q_0$ below which
$C(q)$ is approximately constant. 

\begin{figure}
\includegraphics[width=0.45\textwidth]{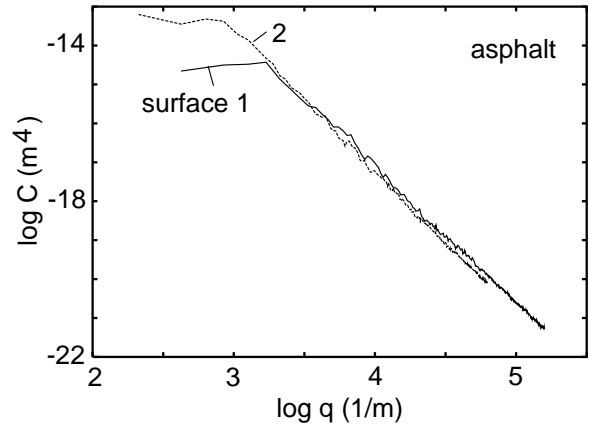}
\caption{\label{CqOpel} 
The 
surface roughness power spectra $C(q)$ for two asphalt road surfaces.} 
\end{figure}

Asphalt and concrete road pavements have nearly perfect self-affine 
fractal power spectra, with very well-defined
roll-off wave vector $q_0 = 2\pi /\lambda_0$ of order 
$1000 \ {\rm m}^{-1}$, corresponding to $\lambda_0 \approx 1 \ {\rm cm}$,
which reflect the largest stone particles used in the asphalt. 
This is illustrated in Fig.~\ref{CqOpel} for two different
asphalt pavements. From the slope of the curves for $q>q_0$ 
one can deduce the fractal dimension $D_{\rm f} \approx 2.2$,
which is typical for asphalt and concrete road surfaces.

\begin{figure}[htb]
   \includegraphics[width=0.3\textwidth]{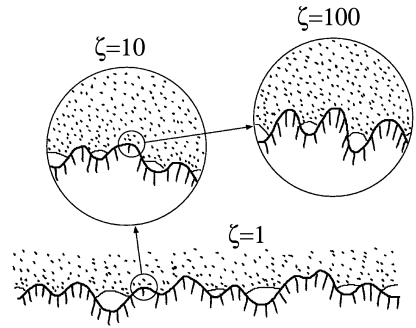}
\caption{
A rubber block (dotted area) in adhesive contact with a hard
rough substrate (dashed area). The substrate has roughness on many different 
length scales and the rubber makes partial contact with the substrate on all length scales. 
When a contact area 
is studied at low magnification ($\zeta=1$) 
it appears as if complete contact occurs in the macro asperity contact regions, 
but when the magnification is increased it is observed that in reality only partial
contact occurs. 
}
\label{1xx}
\end{figure}

Assume that an elastic solid with a flat surface is squeezed against a hard, randomly rough substrate.
Fig. \ref{1xx} shows the 
contact between two solids at increasing magnification $\zeta$. At low magnification
($\zeta \approx 1$)
it looks as if complete contact occurs between the solids at many {\it macro-asperity}
contact regions,
but when the magnification is increased smaller length scale roughness is detected,
and it is observed that only partial contact occurs at the asperities. 
In fact, if there would be no short distance cut-off the true contact area 
would vanish. In reality, however,
a short distance cut-off will always exist since the shortest possible length is 
an atomic distance.

The magnification $\zeta$ refers to some (arbitrarily) chosen reference length scale.
This could be, e.g., the lateral size $L$ of the nominal contact area in which case 
$\zeta= L/\lambda$, where $\lambda$ is the shortest roughness wavelength
components which can be resolved at magnification $\zeta$. In this paper we will 
consider surfaces with power spectra's of the form shown in Fig. \ref{Cq1}, and we will
use the roll-off wavelength $\lambda_0=2\pi /q_0$ as the reference 
length so that $\zeta = \lambda_0/\lambda=q/q_0$.

\begin{figure}[htb]
   \includegraphics[width=0.47\textwidth]{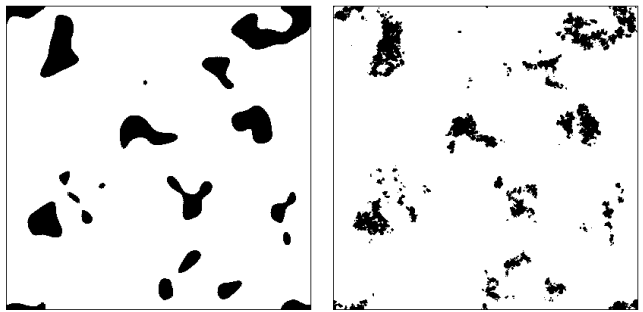}
\caption{
The contact area between an elastic solid with a flat surface and a
hard randomly rough substrate shown at low magnifications
($\zeta = 4$), left, and high magnification ($\zeta=216$), right.
The surface has the fractal dimension $D_{\rm f}=2.2$ and $q_0/q_L = 3$.
Adapted from \cite{Chunyan}.
}
\label{two}
\end{figure}

I now explain the concept of {\it macro-asperity} contact area, which is important for my 
treatment of the influence of the flash-temperature on rubber friction. 
Assume that the surface roughness has the qualitative form shown in Fig. \ref{Cq1} with
a roll-off wave vector $q_0$ corresponding to the magnification $\zeta=1$. In this case 
the macro-asperity contact is the contact region between the solids when the system is studied
at low magnification $\zeta = \zeta_{\rm m} \approx 2-5$ (see Appendix A). 
At this magnification, at low squeezing pressures one observe a 
{\it dilute} distribution of randomly
distributed macro-asperity contact regions with lateral size typically 
of order $\sim \lambda_0/\zeta_{\rm m}$. As the magnification
is increased each macro-asperity contact 
region breaks up into a relative {\it dense} distribution of
much smaller {\it micro-asperity} contact regions. 
This is illustrated in Fig. \ref{two} which shows the result of a
Molecular Dynamics (MD) calculation\cite{Chunyan} where the surface roughness power spectrum 
was assumed to be of
the form shown in Fig. \ref{Cq1} (with the fractal dimension $2.2$).

Consider the contact between two solids at low nominal contact pressure
$\sigma_0 = F_{\rm N}/A_0$, where the contact area
is proportional to the load. Consider first the
system on the length scale $\lambda_0/\zeta_{\rm m}$. 
On this length scale the solids will make (apparent) contact at
a low concentration of (widely separated) contact areas.  
Since the separation between these  macro-asperity
contact regions is very large we can neglect the interaction
between the macro contact regions: in this case  
the pressure in the macro-asperity contact regions 
will be of order $\sim q_0 h_0 E$,
where $h_0$ is the rms roughness amplitude, and $E$ the elastic modulus.
Thus, the (average) pressure in the macroasperity contact
regions is {\it independent} of the nominal contact
pressure $\sigma_0 = F_N/A_0$. Now, each macro-asperity  
is covered by smaller micro asperities, and the smaller  
asperities by even smaller asperities, and so on. 
It is easy to see that at short enough length scale the micro-asperity contact regions will be 
very closely separated and it is therefore impossible to neglect the (elastic or thermal)
interaction between the
asperities at short enough length scale. 

In Sec. 4 we will calculate the flash temperature in the asperity contact regions as a rubber block is
sliding on a rough substrate. We will neglect the thermal coupling 
(or overlap) between the macro-asperity contact regions, 
which
should be a good approximation as long as the contact area at the magnification
$\zeta \approx \zeta_{\rm m}$ is much smaller 
than the nominal contact area (which implies that the distance between
the macro-asperity contact regions is much larger than the 
linear size $\sim \lambda_0/\zeta_{\rm m}$ of these regions). 
However, owing to the high density of 
micro-asperity contact regions within a macro-asperity contact region, it is not possible to
neglect the thermal coupling (or overlap)
between the micro-asperity contact regions within each
macro-asperity asperity contact region. 

\begin{figure}[htb]
   \includegraphics[width=0.40\textwidth]{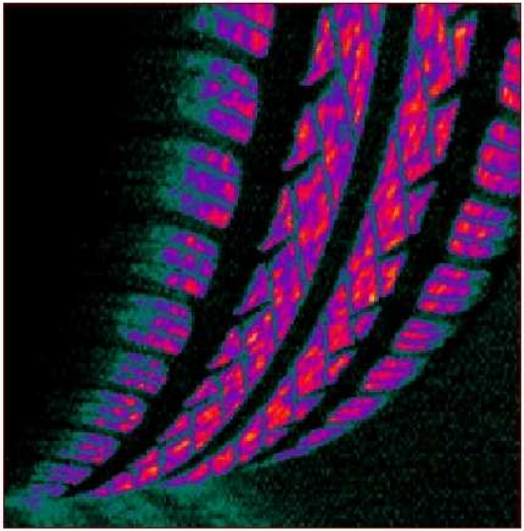}
\caption{
Infrared photo of tire as it rotates out of contact with the road.
The red-yellow color indicates the ``hot'' spots arising from macro-asperity contacts
in the tire-road footprint. Adapted from Ref. \cite{West1,West2}.  
}
\label{infared}
\end{figure}

The temperature increase in the macro-asperity contact regions between a tire and a road surface
has recently been studied using infrared camera.
Fig. \ref{infared} shows a photo of a tire as it rotates out of contact with the road.
The red-yellow color indicate the ``hot'' spots arising from macro-asperity contacts
in the tire-road footprint.

\vskip 0.5cm
{\bf 3 Rubber friction without flash temperature}

\begin{figure}
\includegraphics[width=0.45\textwidth]{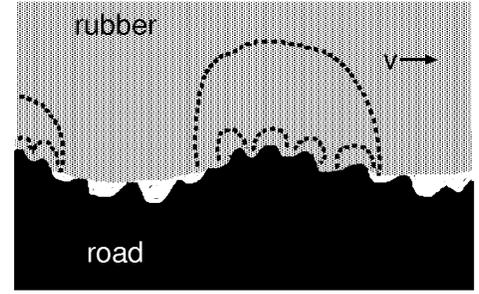}
\caption{\label{deformation} 
The road asperities exert pulsating forces on the sliding rubber block,
leading to energy dissipation in the rubber via the internal friction of the rubber.
Most of the energy dissipation occurs in the volume elements bound by the dashed lines. 
The rubber viscoelastic deformations in the 
large volume elements are induced by the large road asperities, while the smaller dissipative regions 
result from the smaller asperities distributed 
on top of the large asperities. In general, 
in calculating the rubber friction,
the viscoelastic energy dissipation induced
by all the 
asperity sizes must be included, and the local temperature increase (flash temperature) in the
rubber resulting from the energy dissipation must also be taken into account in the analysis.
} 
\end{figure}

The main contribution to rubber friction when a rubber block is sliding on a rough substrate, i.e., a tire
on a road surface, is due to the viscoelastic energy dissipation in the surface region of the rubber as a
result of the pulsating forces acting on the rubber surface from the 
substrate asperities, see Fig.~\ref{deformation}. Recently 
I have developed a theory 
which accurately describes this energy dissipation process, and which predicts the
velocity dependence of the rubber friction
coefficient\cite{PV,P8}. 
The results depend only on the (complex) viscoelastic modulus $E(\omega )$ of the rubber, and on
the substrate surface roughness power spectrum $C(q)$.
Neglecting the flash temperature effect, the kinetic friction coefficient at velocity $v$ is determined
by 
\cite{P8} 
$$\mu_{\rm k} = \frac{1}{2}\int_{q_L}^{q_1} dq \ q^3 C(q) P(q)
\int_0^{2\pi} d\phi \ {\rm cos} \phi \ {\rm Im}
\frac{E(qv \ {\rm cos}\phi )}{(1-\nu^2)\sigma_0},\eqno(3)$$
where 
$$P(q)=  
 {2\over \pi}\int_0^\infty dx \ {{\rm sin} x \over x} {\rm exp}\left (-x^2 G\right ) = 
{\rm erf}\left({1 \over 2\surd G} \right ),\eqno(4)$$
with
$$G(q)=\frac{1}{8}\int_{q_L}^q dq \ q^3C(q)\int_0^{2\pi}d\phi \ \left|\frac{E(qv \ {\rm cos}\phi )}{
(1-\nu^2)\sigma_0}\right|^2,\eqno(5)$$
where $\sigma_0$ is the mean perpendicular pressure (load divided by the 
nominal contact area), and $\nu $ the Poisson ratio, 
which equals 0.5 for rubber-like materials. The factor $P(q)=P(\zeta q_0) = A(\zeta)/A_0$
is the (normalized) area of contact when the system is studied at 
the magnification $\zeta = q/q_0$.

The theory takes into account the substrate surface roughness components
with wave vectors in the range $q_L < q < q_1$, 
where $q_L $ is the smallest relevant wave vector
of order $2\pi /L$, where (in the case of a tire) $L$ is the 
lateral size of a tread block, and where $q_1$ may have 
different origins (see below). Since $q_L$ for a tire tread block is smaller than the 
roll-off wave vector $q_0$ of the
power spectra of most road surfaces (see Fig.~\ref{CqOpel}), rubber friction
is very insensitive to the exact value of $q_L$.

The large wave vector cut off $q_1$ may be related to road contamination, or may be an intrinsic property
of the tire rubber. For example, if the road surface is covered by small contamination particles
(diameter $D$) then $q_1 \approx 2\pi /D$. In this case, 
the physical picture is that when the tire rubber surface is covered by
hard particles of linear size $D$, the rubber will not be able to penetrate 
into surface roughness ``cavities'' with diameter
(or wavelength) smaller than $D$, and such short-range roughness 
will therefore not contribute to the rubber friction.
For perfectly clean road surfaces we believe instead that the 
cut-off $q_1$ is related to the tire rubber properties.
Thus, the high local (flash) temperatures and high local stresses which occur 
in the tire rubber-road asperity contact regions (see below), may
result in a thin (typically of order a few micrometer) 
surface layer of rubber with modified properties (a ``dead'' layer), which may contribute very little
to the observed rubber friction (see Sec. 10). 
Since the stresses and temperatures which develop in the asperity contact regions depend somewhat
on the type of road [via the surface roughness power spectrum $C(q)$], 
the thickness of this ``dead'' layer may vary from
one road surface to another, and some run-in time period will be necessary for a new ``dead'' layer to form when a car
switches from one road surface to another (such ``run-in'' effects are well known experimentally). 

\vskip 0.5cm
{\bf 4 Rubber friction with flash temperature}

\begin{figure}[htb]
\includegraphics[width=0.35\textwidth]{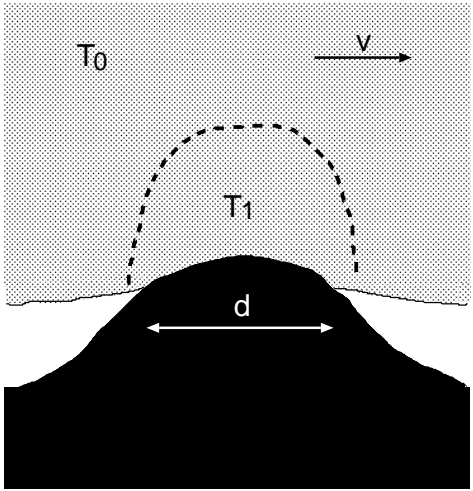}
\caption{A rubber block sliding on a rough hard substrate with surface
roughness on a single length scale. The region in the vicinity of a
substrate asperity is shown.
The heat energy 
production $\dot Q({\bf x},t)$ 
per unit volume and unit time 
occurs mainly within the volume element surrounded by a dashed line.
}\label{Fig.One.Asp} 
\end{figure}

The temperature field $T({\bf x},t)$ in the rubber block is determined by
$${\partial T \over \partial t} -D\nabla^2 T = 
{ \dot Q({\bf x},t)\over \rho C_V}\eqno(6)$$
where $\dot Q$ is the energy production per unit volume and unit time as a result of 
the internal friction in the rubber. The heat diffusivity $D= \lambda / \rho C_V $,
where $\rho$ is the mass density and $\lambda$ the heat conductivity. For rubber
we typically have
$\rho \approx 10^3 \ {\rm kg/m^3}$, $C_V \approx 10^3 \ {\rm J/kg K}$ and
$\lambda \approx 0.1 \ {\rm W/mK}$. This gives $D\approx 10^{-7} \ {\rm m^2/s}$.    
Consider now a rubber block with a flat surface sliding on a rough hard substrate.
Assume that there is an asperity contact area with diameter $d$, 
see Fig. \ref{Fig.One.Asp}. During sliding at the velocity $v$ the asperity will generate 
pulsating forces on the rubber surface
characterized by the frequency $\omega_0 \sim v/d$, which will result in energy dissipation
in a volume element of order $d^3$. If the velocity $v$ is high enough, negligible heat diffusion
will take place during the time period $d/v$ and the temperature increase in the rubber in the vicinity of the
asperity will be $\Delta T \approx Q/\rho C_V$, where $Q$ the the amount of frictional 
heat energy (per unit volume)
in the volume element $d^3$. The assumption of negligible heat diffusion requires that the 
effective contact time $d/v$ is smaller than the diffusion time $d^2/D$, i.e., $vd > D$. 
Thus, for example, if $v=0.1 \ {\rm m/s}$, heat {\it diffusion} will be unimportant 
if $d > D/v \approx 10^{-6} \ {\rm m}$. 
However, after the 
substrate asperity -- rubber contact is ``broken'', 
the peak temperature will decrease, and the temperature distribution
will broaden 
with increasing time, 
in a way determined by the heat diffusion equation.

In this section we will develop a theory for the influence of the flash temperature on rubber
friction when the substrate has roughness on very many different length scales.
We consider first an idealized case where surface roughness occurs on a single length scale,
and then the more complex case of roughness on many different length scales.  
  
We consider a surface with randomness on a single 
length scale $\lambda_0=2\pi/q_0$ and with the
root-mean-square roughness amplitude $h_0$. The surface roughness power spectrum $C(q)=(h_0^2/2\pi q_0)
\delta (q-q_0)$. The (average) radius of curvature
of the asperities is $R\approx 1/(q_0^2h_0)$. Figure \ref{Fig.One.Asp} shows the
contact between one substrate asperity and the rubber. The diameter of the contact area is of order $d$. 
We assume that no 
roughness occurs on smaller length scale than the size of the asperity in Fig. \ref{Fig.One.Asp}.
The time it takes to
slide the distance $d$ is $d/v$. During this time the heat energy produced in the 
volume element $\sim d^3$ at the asperity is 
given by $\mu F_{\rm N} d $, where $F_{\rm N}=\sigma d^2$ is the normal asperity load ($\sigma$
is the average perpendicular stress in the contact area). Thus the energy production per unit
volume 
$Q \approx \mu F_N d /d^3 = \mu \sigma $. Neglecting heat diffusion
this will result in the temperature increase
$\Delta T \approx Q/\rho C_V \approx \mu \sigma /\rho C_V$. 
But we have shown earlier that\cite{PerssonSS}
$$\mu \approx \sigma {{\rm Im}E(\omega_0,T)\over |E(\omega_0,T)|^2}.$$
where $\omega_0=vq_0$.
From standard contact mechanics theory one expect
$\sigma \approx q_0 h_0 |E(\omega_0,T)|$.
Thus the friction 
$$\mu \approx q_0h_0 {{\rm Im}E(\omega_0,T)\over |E(\omega_0,T)|}\eqno(7)$$
and 
the temperature increase
$$\Delta T \approx (q_0h_0)^2  {{\rm Im} E(\omega_0,T)\over \rho C_V}$$
or
$$T \approx T_0 + (q_0h_0)^2 {{\rm Im} E(\omega_0,T)\over \rho C_V}\eqno(8)$$
where $T_0$ is the background temperature.
Note that the complex elastic modulus $E(\omega, T)$
depends on the (local) temperature $T$.
For ``simple'' (unfilled) rubber the {\it Williams-Landel-Ferry equation} (WLF)\cite{WLF}
equation can be used to (approximately) describe the temperature dependence of 
$E(\omega, T)$:
$$E(\omega, T) = E(\omega a_T/a_{T_0}, T_0)\eqno(9a)$$
where
$${\rm log}_{10} a_T \approx -8.86 {T-T_g-50 \over 51.5+T-T_g}.\eqno(9b)$$
For any given viscoelastic modulus $E(\omega, T_0)$, equations (7)-(9)
form a complete set of equations from which the temperature $T$ and the friction coefficient $\mu$ 
can be obtained by, e.g., iteration. Here I will only discuss qualitatively how the
temperature increase $\Delta T$ influence the rubber friction. 

\begin{figure}
\includegraphics[width=0.36\textwidth]{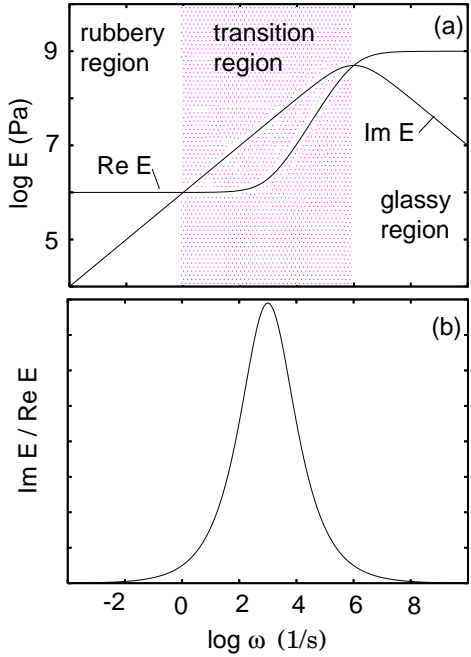}
\caption{\label{newee}
(a) The viscoelastic modulus $E(\omega ) = E_1+iE_2$ of a typical rubber-like
material, and (b) the loss tangent $E_2/E_1$. 
(Schematic.)}     
\end{figure}

In order to understand the qualitative influence of the flash temperature on rubber friction 
it necessary to know the general
structure of the viscoelastic modulus $E(\omega)$ of rubber-like materials. In Fig. \ref{newee}
we show the real $E_1= {\rm Re} E$ and the imaginary part $E_2 = {\rm Im} E$ of $E(\omega)$
and also the loss tangent $E_2/E_1$. At ``low'' frequencies the material is in the ``rubbery'' region
where ${\rm Re} E(\omega)$ is relative small and approximately constant. At very high frequencies
the material is elastically very stiff (brittle-like). In this ``glassy'' region ${\rm Re} E$ is
again nearly constant but much larger (typically by 3 to 4 orders of magnitude) than in the 
rubbery region. In the intermediate frequency range (the ``transition'' region) the loss tangent is very
large and it is mainly this region which determines, e.g., the friction when a tire is sliding on a 
road surface. 

The influence of the flash temperature on rubber friction differs depending on if the
perturbing frequency $\omega_0 = vq_0$ is smaller or larger than the frequency $\omega_1$ where
${\rm Im} E(\omega, T_0)/|E(\omega,T_0)|$ is maximal. 
Since an increase in the temperature ($T_0\rightarrow T >T_0$) shifts
the viscoelastic spectrum to higher frequencies (see the WLF function), 
if $\omega_0 < \omega_1$, 
the flash-temperature
will
decrease the friction. On the other hand, if $\omega_0 > \omega_1$ the opposite effect occur.
However, in most applications, e.g., in tire applications, the perturbing frequencies are 
(almost) always below $\omega_1$
and the friction will decrease when the flash temperature is taken into account. This has extremely important
practical consequences, as we will be discussed later.

We consider now the role of the flash temperature for the general  
(and more complex) case where roughness occurs on many different length scales.
We consider first stationary sliding and then non-stationary sliding.  

\begin{figure}
\includegraphics[width=0.36\textwidth]{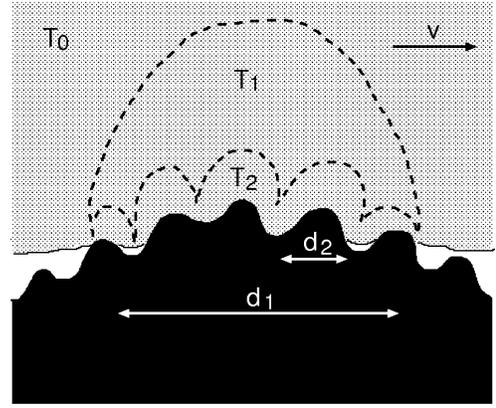}
\caption{\label{bump}
A rubber block sliding on an asperity with shorter wavelength
asperities on top of it. The temperature increases
$T_0 < T_1 < T_2$. 
}     
\end{figure}

\vskip 0.3cm
{\bf 4.1 Stationary sliding}

In this section we will develop a general expression for the friction acting on a rubber
block sliding at a constant velocity on a randomly rough substrate. 
We will 
take into account the effect of the flash temperature. We will including the 
thermal overlap between the heat produced in the micro-asperity contact area
inside every macro-asperity
contact area. That is, 
the temperature rise at one micro-asperity contact area will produce a subsequent
temperature rise at a neighboring micro contact.  
We will make use of a ``mean field'' type of approximation where we {\it laterally} 
smear out the heat
sources associated with the micro-asperity contacts within a macro-asperity contact,
which should be an excellent approximation in most cases. 

Let us now discuss the flash-temperature when a rubber block is sliding on a hard substrate
with surface roughness on many different length scales. The basic problem is illustrated schematically
in Fig. \ref{bump} 
in the case of only two length scales, where a ``large'' asperity is covered by ``small'' asperities.
In response to the large asperity there will be a heating of the rubber in the big volume element
of linear size $d_1$. If $T_0$ denote the background temperature, then the flash 
temperature in the large
volume element will be higher, say $T_1=T_0 +\Delta T_1$. The flash temperature in a small-asperity
volume element (linear size $d_2$, see Fig. \ref{bump}) 
will be even higher, say $T_2 = T_1 +\Delta T_2$. For surfaces with surface roughness on many 
length scales the temperature will increase as we go to 
smaller and smaller asperity contact regions. 
We now study the temperature distribution quantitatively.
 
The average shear stress which acts in the macro asperity-contact area is 
$$\sigma_{\rm m} = \sigma_{\rm f} {A_0 \over A(\zeta_{\rm m})}\eqno(10)$$
where $\sigma_{\rm f}$ is the nominal frictional shear stress, and $A(\zeta_{\rm m})$
the macro asperity contact area observed at the magnification 
$\zeta = \zeta_{\rm m}=\lambda_0/\lambda_{\rm m}
=q_{\rm m}/q_0$ which usually is of order unity (see Appendix A).
Using (3) and (10) and that
$P(q) = A(\zeta )/A_0$ gives
$$\sigma_{\rm m} = {1\over 2}\int d^2q \ q^2 \ {\rm cos} \ \phi \ C(q) {P(q)\over P(q_{\rm m})} 
{\rm Im}{E(qv \ {\rm cos} \ \phi, T_q)\over 1-\nu^2}\eqno(11)$$
where $T_q$ is an effective temperature in the volume of the rubber involved in the calculation
of the contribution to rubber friction from 
surface roughness on the length scale $\lambda= 2 \pi/q$ (see below). 
Thus the energy ``dissipation'' per unit time and unit area in a macro contact area   
$$J = {v\over 2}\int d^2q \ q^2 \ {\rm cos} \ \phi \ C(q) {P(q)\over P(q_{\rm m})} 
{\rm Im}{E(qv \ {\rm cos} \ \phi,T_q)\over 1-\nu^2}\eqno(12)$$
The energy production per unit volume, from asperities on the
length scale $\lambda = 2\pi /q$, decay into the solid as $\sim {\rm e}^{-2qz}$ so that
the energy production per unit volume and unit time in a contact area is
obtained by introducing in the integrand in (12) the factor
$${{\rm e}^{-2qz} \over \int_0^\infty
dz \ {\rm e}^{-2qz}}=
2q \ {\rm e}^{-2qz}\eqno(13)$$
Using this equation 
the energy production per unit volume and unit time 
in the rubber in a macro-asperity contact area is given by:
$$\dot Q(z,t) = \theta (t) {v\over 2}\int d^2q \ 2q^3 {\rm e}^{-2qz} 
\ {\rm cos} \ \phi \ C(q)$$
$$\times {P(q)\over P(q_{\rm m})} 
{\rm Im}{E(qv \ {\rm cos} \ \phi, T_q)\over 1-\nu^2}\eqno(14)$$
where we have assumed that the energy production start at $t=0$.
We write the step
function $\theta (t)$ as 
$$\theta (t) = -{1\over 2 \pi i}\int d\omega \ {1\over \omega + i \epsilon} {\rm e}^{-i\omega t}$$
where $\epsilon$ is a positive infinitesimal number.
We can also write
$${\rm e}^{-2q|z|} = {1\over 2 \pi} \int dk \ {4q \over k^2 + 4q^2} \ {\rm e}^{ikz}$$  
Thus
$${\dot Q(z,t) \over \rho C_V} = \int_0^{\infty} dq \ f(q)$$ 
$$\times {(-1)\over i (2\pi)^2 } \int d\omega dk 
 \ {1\over \omega + i \epsilon} 
 \ {4q \over k^2 + 4q^2} 
{\rm e}^{i(kz-\omega t)}\eqno(15)$$
where
$$f(q)= 
{v q^4 \over \rho C_V} C(q) {P(q)\over P(q_{\rm m})} 
\int d \phi \ {\rm cos} \phi \ {\rm Im}{E(qv \ {\rm cos} \ \phi,T_q)\over 1-\nu^2}\eqno(16)$$
To solve the heat diffusion equation (6) we need a boundary condition on the
surface $z=0$. We will assume that $\partial T /\partial z = 0$ for $z=0$, i.e.,
we neglect heat transfer from the rubber to the substrate. 
This is an excellent approximation at high enough sliding velocity. At low sliding velocity it will
give rise to an overestimation of the flash temperature, but in this case the 
flash temperature effect is anyhow not very important.
The boundary condition $\partial T/\partial z (z=0)=0$ is equivalent to solving
the heat diffusion equation for an extended solid ($-\infty < z < \infty$) with a symmetric
heat source obtained by replacing ${\rm exp}(-2qz)$ in (14) by ${\rm exp}(-2q|z|)$. Thus we get
$$T(z,t) = T_0+ 
\int_0^{\infty} dq \ f(q)$$ 
$$\times {(-1)\over (2\pi)^2 } \int d\omega dk \ {1\over \omega + i Dk^2}
 \ {1\over \omega + i \epsilon} 
 \ {4q \over k^2 + 4q^2} 
{\rm e}^{i(kz-\omega t)}\eqno(17)$$
Performing the $\omega$-integration gives
$$T(z,t) = T_0+
\int_0^{\infty} dq \ f(q)$$ 
$$ \times {1\over 2 \pi} \int dk \ {1\over Dk^2} \left (1-{\rm e}^{-Dk^2t}\right ) 
 \ {4q \over k^2 + 4q^2} 
{\rm e}^{ikz}\eqno(18)$$
The optimum (spatially uniform) 
temperature to be used when calculating the contribution to the friction
from surface roughness on the length 
scale $\lambda =2 \pi /q$ can be obtained using
$$T_{q} = {\int_0^\infty dz \ T(z,t) {\rm e}^{-2qz} \over
\int_0^\infty dz \ {\rm e}^{-2qz}}\eqno(19)$$ 
Using (18) this gives
$$T_{q} = T_0+
\int_0^{\infty} dq' \ g(q,q') f(q')\eqno(20)$$ 
where
$$g(q,q') = 
{1\over \pi} \int_0^{\infty} dk \ {1\over Dk^2} \left (1-{\rm e}^{-Dk^2t_0}\right )$$
$$\times 
{4q' \over k^2 + 4q'^2} \  
{4q^2 \over k^2+ 4q^2}\eqno(21)$$
where $t_0 \approx R/v$ is roughly half the time a rubber patch is in contact with the
macro asperity. The radius $R$ of a macro-asperity contact region is estimated in Appendix A.
Note that the complex elastic modulus $E(\omega, T)$
depends on the (local) temperature $T$.
In general one may write $E(\omega,T)=b_TE(\omega a_T, T_0)$, where $a_T$ and $b_T$ depend on the temperature
$T$ but with $a_{T_0}=b_{T_0}=1$. For unfilled rubber $b_T\approx 1$ and $a_T$ is (approximately) given by the
WLF equation.  
The functions $a_T$ and $b_T$ are today measured routinely using standard
rheological equipment.

When $T_q$ has been obtained from (20) and (21), the friction coefficient can be calculated using the
equations derived in Ref. \cite{P8}: 
$$\mu \approx {1\over 2} \int_{q_0}^{q_1}  dq \ q^3
 \ C(q) P(q)$$
$$\times \int_0^{2\pi} d\phi \ {\rm cos}\ \phi \ 
{\rm Im}{E(qv \ {\rm cos} \ \phi, T_q)\over (1-\nu^2)\sigma_0}$$
$$P(q)={2\over \pi} \int_0^{\infty} dx \ {{\rm sin}x \over x} 
{\rm exp} \left [-x^2G(q)\right ]$$
where
$$G(q)={1\over 8}\int_{q_0}^q dq \ q^3 C(q) 
\int_0^{2\pi} d\phi \  
\left | {E(qv \ {\rm cos} \ \phi, T_q)\over (1-\nu^2)\sigma_0}\right | ^2$$

\vskip 0.3cm
{\bf 4.2 Non-stationary sliding}

Here we develop a completely general theory of non-stationary sliding. 
Consider again the heat diffusion equation 
$${\partial T \over \partial t} -D\nabla^2 T = 
{\dot Q(z,t)\over \rho C_V} \theta \left (R-|{\bf x}-{\bf r}(t)| \right )\eqno(22)$$
where ${\bf r}(t) = \hat x x(t)$ is the position vector of 
the bottom surface of the rubber block, and where $R$ is the radius of a macro-asperity contact region
which we assume to be circular. Using (14) and (16) gives
$${\dot Q(z,t)\over \rho C_V}=\int_0^\infty dq \ f(q,t) \ e^{-2qz}.\eqno(23)$$
Let us introduce a new coordinate system, moving with the bottom surface of the rubber block:
$${\bf x}' = {\bf x}-{\bf r}(t)$$
Substituting this in (22), and replacing ${\bf x}' \rightarrow {\bf x}$ for simplicity, gives
$${\partial T \over \partial t} -\dot {\bf r}(t)\cdot {\partial T\over \partial {\bf x}}
-D{\partial^2 T\over \partial z^2} = 
{\dot Q(z,t)\over \rho C_V} \theta \left (R-|{\bf x}| \right )\eqno(24)$$
We now introduce the Fourier transform
$$\tilde T({\bf p},k,t)={1\over (2\pi)^3} \int d^2x dz \ T({\bf x},z,t) 
e^{-i({\bf p}\cdot {\bf x}+kz)}\eqno(25)$$
$$T({\bf x},z,t)=\int d^2p dk \ \tilde T({\bf p},k,t) 
e^{i({\bf p}\cdot {\bf x}+kz)}\eqno(26)$$
Using (25) we get from (24):
$${\partial \tilde T \over \partial t} -i{\bf p}\cdot {\dot {\bf r}(t)}\tilde T
+Dk^2 \tilde T$$
$$ = 
{\dot Q(k,t)\over \rho C_V} {1\over 2 \pi}\int_0^R dr r J_0(kr)\eqno(27)$$ 
where $J_0(x)$ is the zero order Bessel function and where
$$\dot Q(k,t) = {1\over 2 \pi} \int dz \ \dot Q(z,t) e^{-ikz}.\eqno(28)$$
Equation (27) is easy to integrate to get
$$\tilde T({\bf p},k,t)
=  
\int_0^t dt' \ e^{-Dk^2(t-t')+ip_x [x(t)-x(t')]}$$
$$\times 
{\dot Q(k,t')\over \rho C_V} {1\over 2 \pi}\int_0^R dr r J_0(kr)\eqno(29)$$
The laterally averaged temperature in the contact area is
$$\bar T(k,t) = {1\over \pi R^2} \int_{|{\bf x}| <R} d^2x \ T({\bf x},k,t)$$ 
$$= {1\over \pi R^2} \int_{|{\bf x}| <R} d^2x \int d^2p \ e^{i{\bf p}\cdot {\bf x}}
\tilde T({\bf p},k,t)$$ 
$$= {2\over R^2} \int d^2p \int_0^R dr r J_0(kr)\tilde T({\bf p},k,t)\eqno(30)$$
Substituting (29) in (30) gives after some simplifications
$$\bar T(k,t) = 
\int_0^t dt' \ \Gamma (t,t') e^{-Dk^2(t-t')}
{\dot Q(k,t')\over \rho C_V}\eqno(31)$$
where
$$\Gamma (t,t')= {2\over R^2} \int_0^\infty dp p 
J_0(p[x(t)-x(t')])$$
$$\times \left ( \int_0^R dr r J_0(pr)\right )^2$$
Now since
$$\int_0^R dr r J_0(pr) = {R\over p}J_1(pR),$$
we get with $pR=y$
$$\Gamma (t,t')= 2 \int_0^\infty dy y^{-1} 
J_0(2wy)\left (J_1(y)\right )^2=h(w)\eqno(32)$$
where
$$w=w(t,t')=[x(t)-x(t')]/2R\eqno(33)$$
Performing the integral in (32) gives
$$h(w)=1-{2\over \pi}w \left (1-w^2\right )^{1/2}-{2\over \pi}
{\rm arcsin}(w)\eqno(34)$$  
for $w<1$ while $h=0$ for $w>1$. 
Using (23) we get
$${\dot Q(k,t)\over \rho C_V} = {1\over 2 \pi}\int_0^\infty dq \ f(q,t) 
{4q\over k^2+4q^2}\eqno(35)$$
Next, using that
$$\bar T_q (t) = 2q\int_0^\infty dz \ \bar T(z,t) e^{-2qz}$$
$$=2 \int_0^\infty dk \ {4q^2\over k^2+4q^2}\bar T(k,t)\eqno(36)$$
we get from (31), (32), (35) and (36)
$$\bar T_q (t) = T_0+\int_0^t dt' \ \Gamma (t,t')\int_0^\infty dq' f(q',t')$$
$$\times {1\over \pi} \int_0^\infty dk \ {4q^2\over k^2+4q^2} \ {4q'\over k^2+4q'^2}
e^{-Dk^2(t-t')}\eqno(37)$$

Let us summarize the basic equations derived above. 
The friction coefficient
$$\mu (t) \approx {1\over 2} \int_{q_0}^{q_1}  dq \ q^3
 \ C(q) P(q,t)$$
$$\times \int_0^{2\pi} d\phi \ {\rm cos}\ \phi \ 
{\rm Im}{E(qv(t) \ {\rm cos} \ \phi, T_q(t))\over (1-\nu^2)\sigma_0}\eqno(38)$$
where $\sigma_0 = F_N(t)/A_0$, where $A_0$ is the nominal contact area.
In this equation enters the flash temperature at time $t$:
$$T_q (t) = T_0+\int_0^t dt' \ \Gamma (t,t')\int_0^\infty dq' f(q',t')$$
$$\times {1\over \pi} \int_0^\infty dk \ {4q^2\over k^2+4q^2} \ {4q'\over k^2+4q'^2}
e^{-Dk^2(t-t')}
\eqno(39)$$
where $\Gamma (t,t')=h(w(t,t'))$ with $w(t,t')=[x(t)-x(t')]/2R$ depend on the history of the sliding motion.
The function        
$$f(q,t)= 
{v(t) \over \rho C_v} q^4 C(q) {P(q,t)\over P(q_{\rm m},t)}$$ 
$$\times \int d \phi \ {\rm cos} \phi \ {\rm Im}{E(qv(t) \ {\rm cos} \ \phi,T_q(t))\over 1-\nu^2},\eqno(40)$$
where $v=\dot x(t)$ depends on time. The function $P(q,t)$ (which also depends on time) is given
by the standard formula:
$$P(q,t)={2\over \pi} \int_0^{\infty} dx \ {{\rm sin}x \over x} 
{\rm exp} \left [-x^2G(q,t)\right ]
={\rm erf}\left ({ 1\over 2 \surd G}\right )\eqno(41)$$
where
$$G(q,t)={1\over 8}\int_{q_0}^q dq \ q^3 C(q) 
\int_0^{2\pi} d\phi \  
\left | {E(qv(t) \ {\rm cos} \ \phi, T_q(t))\over (1-\nu^2)\sigma_0}\right | ^2\eqno(42)$$

\vskip 0.3cm
{\bf 4.2.1 Limiting cases and physical interpretation of $h(w)$} 

When the heat diffusivity $D=0$ equation (39) takes the form
$$\bar T_q (t) = T_0+\int_0^t dt' \ \Gamma (t,t')\int_0^\infty dq' f(q',t')
{q \over q+q'}\eqno(43)$$
In the limit $D\rightarrow \infty$ we get as expected $T=T_0$. 

Finally, let us consider stationary sliding so that $x(t)=v_0t$. In this case
$w=(v_0/2R)(t-t')$ and $f(q,t)=f(q)$ is independent of time. Thus (39) takes the form       
$$\bar T_q (t) \approx T_0+\int_0^\infty dq' G(q,q') f(q')\eqno(44)$$
where
$$G(q,q') =
{1\over \pi} \int_0^\infty dk \ {4q^2\over k^2+4q^2} \ {4q'\over k^2+4q'^2} M(k)\eqno(45)$$
where
$$M(k)=\int_{t-2R/v_0}^t dt' \ h(w) e^{-Dk^2(t-t')}\eqno(46)$$
Changing the integration variable from $t'$ to $w$ gives 
$$M(k)= {2R\over v_0} \int_{0}^1 dw \ h(w) e^{-(2R/v_0) Dk^2 w }\eqno(47)$$
The result (45) is the same as (21) but with a slightly different (and more accurate) function $M(k)$.
Thus, in (21) occur instead of $M(k)$ the factor
$${1\over Dk^2} \left (1- e^{-Dk^2t_0}\right ) = t_0 \int_0^1 dw \ e^{-t_0Dk^2 w}$$
Since $t_0$ is of order $R/v_0$ and $h(w)$ of order 1, these two expressions for the steady
rubber friction are of similar magnitude.

\begin{figure}[htb]
\includegraphics[width=0.3\textwidth]{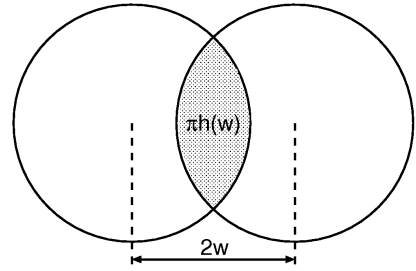}
\caption{The common region (dotted area)
between the two circles has the area $\pi h(w)$.} 
\label{TwoC}
\end{figure}

Here I present an alternative derivation of the
factor $h(w)=g(t,t')$ which gives a direct 
physical interpretation of this factor. In the theory above 
this function emerged purely mathematically, as a
result of an integral of a product of three Bessel functions.

Let us assume that the heat diffusivity $D=0$ and consider
the equation
$${\partial T\over \partial t} = K(z,t)\theta(R-|{\bf x}-{\bf r}(t)|)\eqno(48)$$
where $K=\dot Q/\rho C_V$ and where $\theta(x)$ is the step function. Here $R$ is the
radius of the macro contact area and ${\bf r}(t)$ the position vector of the contact area.
Integrating (48) gives
$$T({\bf x},z,t)=T_0+\int_0^t dt' \  
\theta(R-|{\bf x}-{\bf r}(t')|)K(z,t')\eqno(49)$$
Averaging the temperature (49) over the contact area gives
$$\bar T(z,t) = 
T_0+{1\over \pi R^2}\int_{|{\bf x}-{\bf r}(t)|<R} d^2x$$
$$\times \int_0^t dt' \  
\theta(R-|{\bf x}-{\bf r}(t')|)K(z,t')$$
$$=T_0+{1\over \pi R^2}\int_0^t dt' \ 
K(z,t')$$
$$\times \int_{|{\bf x}|<R} d^2x \   
\theta(R-|{\bf x}+{\bf r}(t)-{\bf r}(t')|)\eqno(50)$$
If we change integration variable ${\bf x}' = {\bf x}/R$ and denote ${\bf x}'$ with ${\bf x}$
for simplicity, we get from (50)
$$\bar T =T_0+{1\over \pi }\int_0^t dt' \ K(z,t') \int_{|{\bf x}|<1} d^2x  \  
\theta(1-|{\bf x}+2{\bf w}|)\eqno(51)$$
where
$${\bf w} =
[{\bf r}(t)-{\bf r}(t')]/2R$$
Note that the integral
$$h(w) = 
{1\over \pi} \int_{|{\bf x}|<1} d^2x \   
\theta(1-|{\bf x}+2{\bf w}|)\eqno(52)$$
is the common area (divided by $\pi$) between two circular regions (with unit radius) with the origins
separated by the distance $2w$, see Fig. \ref{TwoC}. It is easy to calculate
$$h(w)=1-{2\over \pi}w \left (1-w^2\right )^{1/2}-{2\over \pi}
{\rm arcsin}(w)\eqno(53)$$  
for $w<1$ while $h(w)=0$ for $w>1$. 
This formula agrees with (34). Note that $w=w(t,t')$ and $h(w)=\Gamma (t,t')$ so that (51)-(53) gives 
$$\bar T =T_0+{1\over \pi }\int_0^t dt' \ \Gamma (t,t') K(z,t') \eqno(54)$$
Finally, using (14), (16) and (19) in (54) gives
$$\bar T_q (t) = T_0+\int_0^t dt' \ \Gamma (t,t') \int_0^\infty dq' f(q',t')
{q\over q+q'}$$
which is the same as (43). Thus, we conclude that the factor
$h(w)=\Gamma (t,t')$ in (37), is equal to the overlap area (divided by $\pi$)
between the contact area at time $t$, and the
contact area at an earlier time $t'$, see Fig. \ref{TwoC}.

\begin{figure}[htb]
\includegraphics[width=0.40\textwidth]{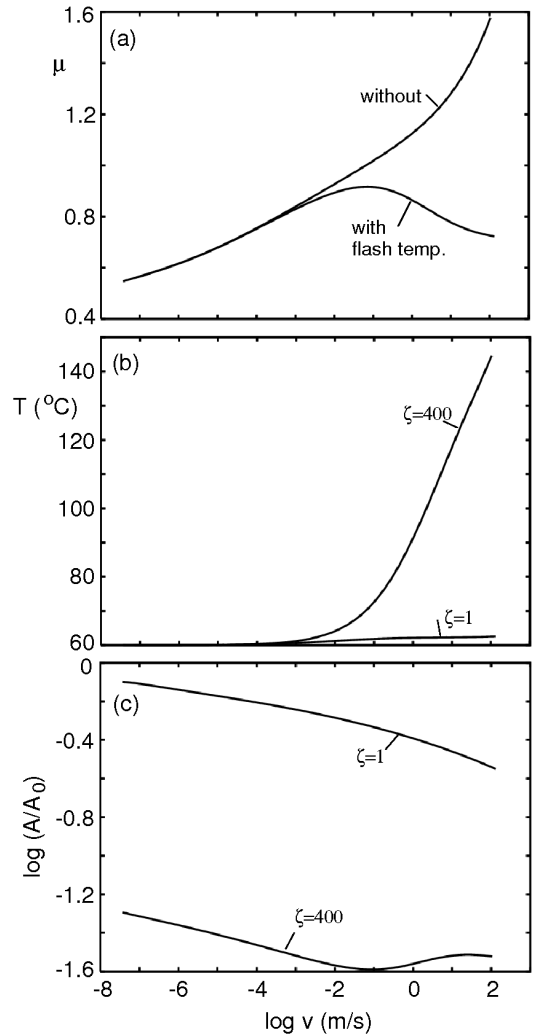}
\caption{
The friction coefficient (a), the flash temperature (b), 
and the logarithm of the (normalized) area of contact (c),
as a function of the logarithm of the sliding velocity.
For a rubber block (tire tread rubber) sliding on an asphalt road surface.
The background temperature $T_0=60 \ ^\circ C$ and $\zeta_{\rm max}=400$.
In (a) we show the kinetic friction coefficient both with and without the flash temperature.
In (b) and (c) we show results both for $\zeta=1$ and at the highest magnification $\zeta=400$.} 
\label{mtav}
\end{figure}

\vskip 0.5cm
{\bf 5 Sliding dynamics: numerical results}

In this section
we will present numerical results for the friction force acting on a 
rubber (tire tread) block, sliding on an asphalt road surface.
We will study the friction as a function of the velocity of the {\it bottom} surface of the block.
In real experiments it is, of course, not possible to specifies the motion of the bottom surface
directly (unless steady sliding occur), but usually one specify the motion of the top surface of the rubber 
block (or some other distant
part of the system). We will consider this 
case later.
In this section we assume the rubber background temperature $T_0=60 \ ^\circ {\rm C}$,
the substrate is an asphalt road surface (with the power spectrum given in Fig. \ref{CqOpel})
and include the substrate
roughness down to the wave vector cut-off $q_1 = \zeta_{\rm max} q_0$ with $\zeta_{\rm max} = 400$ and
$q_0 = 1500 \ {\rm m}^{-1}$  so that $q_1 = 6\times 10^5$ or 
$\lambda_1 = 2 \pi /q_1 \approx 10 \ {\rm \mu m}$. 
All calculations is for the nominal (squeezing) pressure $\sigma_0=0.4 \ {\rm MPa}$.

\begin{figure}[htb]
\includegraphics[width=0.40\textwidth]{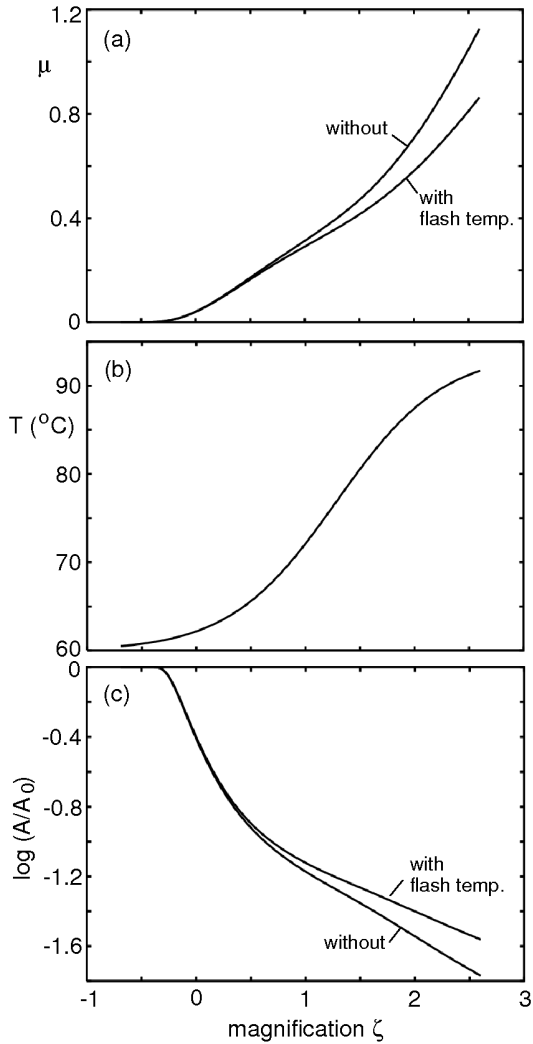}
\caption{
The friction coefficient (a), the flash temperature (b), 
and the logarithm of the (normalized) area of contact (c),
as a function of the logarithm of the magnification.
For a rubber block (tire tread rubber) 
sliding on an asphalt road surface.
The background temperature $T_0=60 \ ^\circ C$ and the sliding velocity $v= 1 \ {\rm m/s}$.
In (a) and (c) we show 
results both with and without the flash temperature.
} 
\label{mtaz}
\end{figure}

\vskip 0.3cm
{\bf 5.1 Stationary motion}

Fig. \ref{mtav} shows 
the friction coefficient (a), the flash temperature (b), 
and the logarithm of the (normalized) area of contact (c),
as a function of the logarithm of the sliding velocity.
In (a) we show the kinetic friction coefficient 
both with and without the flash temperature.
In (b) and (c) we show results both for $\zeta=1$ and 
at the highest magnification $\zeta=400$. 
Note that without the flash temperature the rubber friction coefficient increases monotonically
up to the highest studied sliding velocity $v \approx 100 \ {\rm m/s}$. When the 
flash temperature is included in the study the friction is maximal for $v\approx 1 \ {\rm cm/s}$.
The decrease of the friction observed when the flash temperature is included in the analysis 
is easy to understand (see also Sec. 4): 
when the rubber heats up the viscoelastic modulus $E(\omega )$ shift towards higher 
frequencies and the rubber becomes more elastic 
(less viscous) resulting in less energy dissipation. 

Fig.  \ref{mtav}(b) shows that the flash temperature at the magnification $\zeta = 400$
(i.e., the temperature about 
$10 \ {\rm \mu m}$ below the rubber surface) 
is $\approx 140 \ ^\circ {\rm C}$, i.e.,
about $80 \ ^\circ {\rm C}$  above the rubber background temperature.
On the other hand the temperature increase a few mm below the surface (corresponding to the magnification
$\zeta=1$) is just a few degree.
We note that during steady sliding the temperature in the 
whole rubber block will increase continuously with increase time, but this effect is not included in
the discussion above, unless one allowed for the background temperature 
$T_0$ to increase with increasing time. The calculation of the time dependence 
of $T_0$ require the knowledge of the temperature of the surrounding medium, and the heat transfer 
coefficient to the surrounding. This topic is important in many practical
applications, but does not interest us here.

Fig.  \ref{mtav}(c) shows that the 
area of (apparent) contact at low magnification ($\zeta=1$) decreases continuously with increasing
sliding velocity. This result is expected because the frequencies $\omega$ 
of the pulsating deformations of the rubber
increases with increasing sliding velocity $v$ ($\omega \sim v$), and rubber becomes elastically 
stiffer at higher frequencies, resulting in an (apparent) smaller contact area at high
sliding velocity. However, at the highest
magnification $\zeta = 400$, fig. \ref{mtav}(c) shows that the area of contact {\it increases} with 
increasing sliding velocity for $v > 0.1 \ {\rm m/s}$. This is caused by the increase in the temperature
in the surface region of the rubber, which shift the viscoelastic modulus towards higher frequencies faster than
the linear increase $\omega \sim v$ in the perturbing frequencies arising from the increase in the sliding 
velocity.

Fig. \ref{mtaz}
shows (a) the friction coefficient, (b) the flash temperature, 
and (c) the logarithm of the (normalized) area of contact,
as a function of the logarithm of the magnification,
for the sliding velocity $v= 1 \ {\rm m/s}$.
In (a) and (c) we show 
results both with and without the flash temperature.
Fig. \ref{mtaz}(a) shows that when the flash temperature is included,
the surface roughness on every decade in length scale contributes roughly with an equal amount to the rubber
sliding friction. Fig. \ref{mtaz}(c) shows that at low magnification the contact area
is the same with and without the flash temperature included in the analysis, while for
large magnification the area of contact increases when the flash temperature is included, as expected from the
elastic softening of the rubber at higher temperatures (see above).

\begin{figure}[htb]
\includegraphics[width=0.40\textwidth]{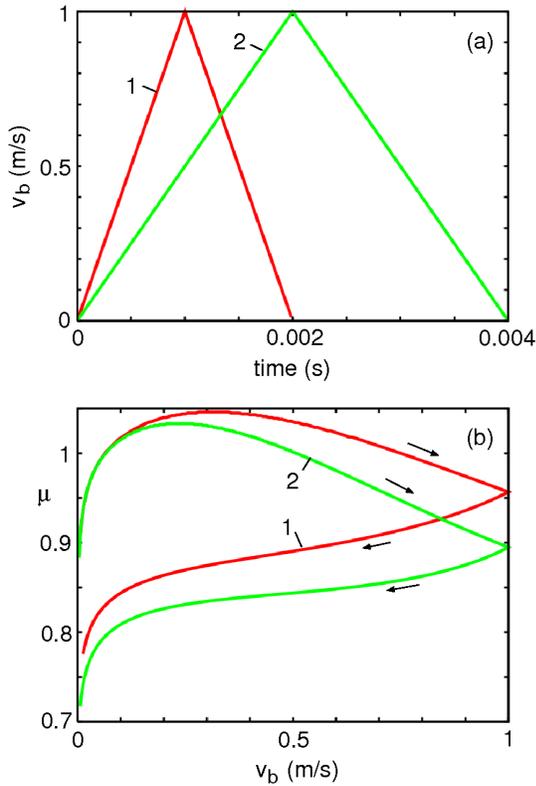}
\caption{
Non-uniform sliding motion. (a) The velocity of the bottom surface
of the block increases linearly with time from zero to $1 \ {\rm m/s}$
and then decreases back to zero. In case {\bf 1} the time for the
whole velocity cycle is $0.002 \ {\rm s}$ and for case {\bf 2}, $0.004 \ {\rm s}$.
(b) The friction coefficient exhibit hysteresis as a function of the sliding velocity,
and is smaller during the retardation
time period than during the acceleration time period. This behavior reflect the build up of the
flash temperature during the sliding cycle.
} 
\label{mva}
\end{figure}

\begin{figure}[htb]
\includegraphics[width=0.4\textwidth]{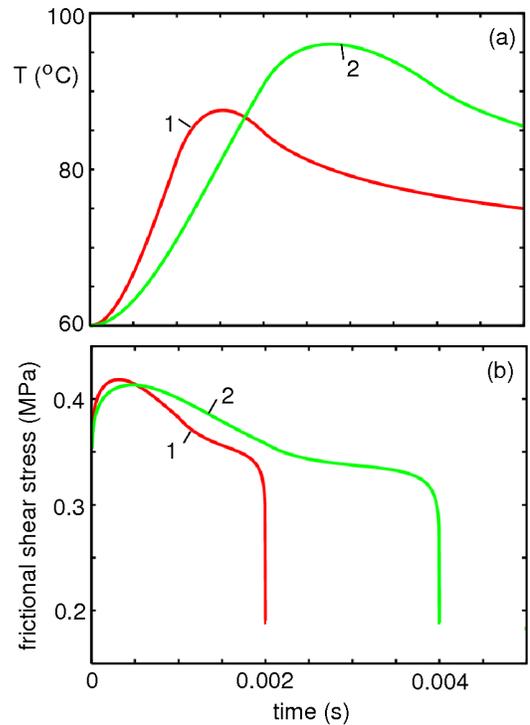}
\caption{
The flash temperature (a) and the frictional shear stress (b) as a function of time
as the bottom surface of the rubber block perform the motion indicated in Fig. \ref{mva}(a),
and with $v_{\rm b}=0$ for $t>0.002 \ {\rm s}$ and $0.004 \ {\rm s}$ for case {\bf 1} and
{\bf 2}, respectively. 
} 
\label{T.stress}
\end{figure}

\vskip 0.3cm
{\bf 5.2 Non-stationary motion}

It is usually assumed in simple treatments of friction, that the friction
coefficient is a function only of the {\it instantaneous} slip velocity $v_{\rm b}(t)$,
i.e., $\mu(t)= \mu (v_{\rm b}(t))$, where $v_{\rm b}(t)$ is the velocity of the {\it bottom} 
surface (in contact with the substrate) of
the sliding block. However, this approximation fails badly for rubber friction.
The reason is the strong dependence of rubber friction on the temperature distribution 
in the rubber block. Since the temperature distribution $T({\bf x}, t)$ at time $t$ depends on the
sliding motion for all earlier times $t' \le t$, it follows immediately that
for rubber-like materials the fiction coefficient will depend on the 
{\it sliding history}, i.e., it will be a functional of $v_{\rm b}(t)$: 
$\mu = \mu (v_{\rm b}(t'), t' \le t)$. In this section I present numerical results which
illustrate this fact.

\begin{figure}[htb]
\includegraphics[width=0.4\textwidth]{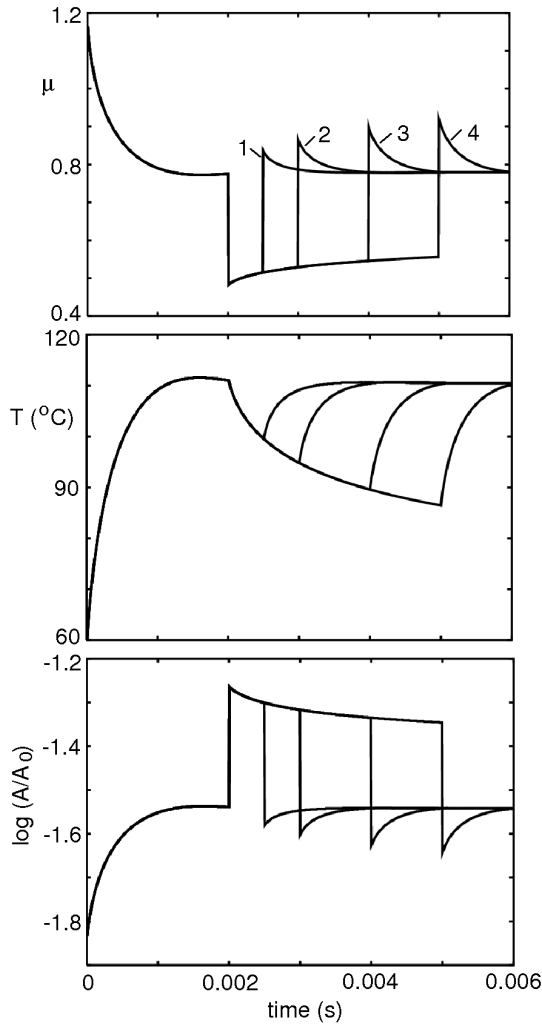}
\caption{
(a) the friction coefficient, (b) the flash-temperature at the highest magnification and (c)
the (relative) contact area at the highest magnification, as a function of time. The bottom
surface of the block moves with the constant velocity $v_{\rm b} = 2 \ {\rm m/s}$ 
for $0 < t < 0.002 \ {\rm s}$, and 
at $t=0.002 \ {\rm s}$ it is abruptly reduced to $10^{-5} \ {\rm m/s}$ and is kept 
at this velocity for 
$0.0005 \ {\rm s} $ (case {\bf 1}),
$0.001 \ {\rm s} $ (case {\bf 2}),
$0.002 \ {\rm s} $ (case {\bf 3}),
$0.003 \ {\rm s} $ (case {\bf 4}), and then returned to $v_{\rm b}=2 \ {\rm m/s}$.
} 
\label{muTrelA}
\end{figure}

Let us assume the velocity of the bottom surface
of the block first increases linearly with time from zero to $1 \ {\rm m/s}$
and then decreases back to zero, as indicated in Fig. \ref{mva}(a). 
In case {\bf 1} the time for the
whole cycle is $0.002 \ {\rm s}$ and for case {\bf 2}, $0.004 \ {\rm s}$.
In Fig. \ref{mva}(b) we show the resulting friction coefficient. 
Note that $\mu$ exhibits hysteresis as a function of the sliding velocity,
and that $\mu$ is smaller during the retardation than during the acceleration time period. 
This behavior just reflects the finite slip-distance necessary for building up the
flash temperature, and the finite time involved in heat diffusion, 
and shows that $\mu$ does not just depend on the
instantaneous sliding velocity but on the whole sliding history (memory-effects).
Note that the slip-distance is given by $v_{\rm max}^2/a$, where $v_{\rm max}= 1 \ {\rm m/s}$
is the maximum of the velocity and the acceleration $a=1000$ and $500 \ {\rm m/s^2}$ for case
{\bf 1} and {\bf 2} respectively, giving the slip distances 1 and 2 mm, respectively.
These values are both smaller than the diameter of the macro-asperity
contact regions which is $2R\approx 3.3 \ {\rm mm}$.   
The larger slip distance for case {\bf 2} implies that more energy is deposited (in the rubber) 
at the rubber-substrate contact regions than for case {\bf 1}, 
which will result in a higher flash temperature in case
{\bf 2} than in case {\bf 1}. This explains why the friction coefficient 
is smaller for case {\bf 2} (see Fig. \ref{mva}(b)).

The temperature buildup during the sliding motion is illustrated in Fig. \ref{T.stress}.
This figure also shows that after the sliding motion has stopped, 
i.e., for $t>0.002 \ {\rm s}$ and $t>0.004 \ {\rm s}$ for case {\bf 1} and {\bf 2}, respectively,
the temperature at the surface
decreases monotonically because of heat diffusion. 
Note that as a function of the slip velocity $v_{\rm b}$ (not shown), 
the flash temperature in Fig.  \ref{T.stress}(a) 
is always higher for case {\bf 2} as compared to case {\bf 1}.
Fig. \ref{T.stress}(b) shows the frictional shear stress 
$\sigma_{\rm f}(t) =\sigma_0 \mu (t)$ as a function of time.

Assume now that the bottom
surface of the block moves with the constant velocity $v_{\rm b} = 2 \ {\rm m/s}$ 
for $0 < t < 0.002 \ {\rm s}$, and 
at $t=0.002 \ {\rm s}$ it is abruptly reduced to $10^{-5} \ {\rm m/s}$, and is then kept 
at this value
for 
$0.0005 \ {\rm s} $ (case {\bf 1}),
$0.001 \ {\rm s} $ (case {\bf 2}),
$0.002 \ {\rm s} $ (case {\bf 3}),
$0.003 \ {\rm s} $ (case {\bf 4}), and then returned to $v_{\rm b}=2 \ {\rm m/s}$.
For this sliding history we show in Fig. \ref{muTrelA}  
(a) the friction coefficient, (b) the flash-temperature at the highest magnification and (c)
the (relative) contact area at the highest magnification, as a function of time. 
Note that the longer the system is kept in the {\it low-velocity state} (where
$v=10^{-5} \ {\rm m/s}$), the higher the
``start-up'' peak in the friction will be when the velocity is switched back to
$2 \ {\rm m/s}$. The reason for this is that, because of heat diffusion, in
the low-velocity state the temperature 
in 
the rubber 
will decrease continuously with increasing time,
see Fig. \ref{muTrelA}(b), and the rubber friction increases when 
the rubber temperature decreases.
Fig. \ref{muTrelA}(c) illustrate that, as expected, the contact area between the rubber 
and the substrate increases 
when the rubber heats up and when the
rubber sliding velocity decreases.

\begin{figure}[htb]
\includegraphics[width=0.4\textwidth]{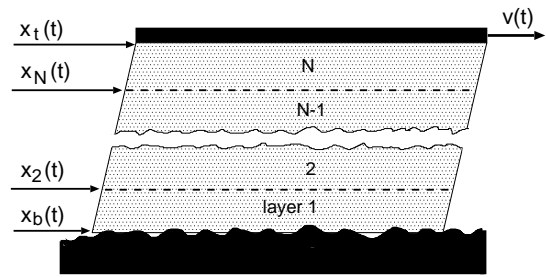}
\caption{
A rubber block (thickness $d$) pulled on a rough hard substrate. 
The upper surface is clamped and moved with a prescribed velocity $v(t)=\dot x_{\rm t}$.
In the mathematical description of the sliding motion, the block is discretized
into $N$ layers of thickness $\Delta d = d/N$. The coordinates $x_1(t)= x_{\rm b}(t)$,
$x_2(t), ..., x_{\rm N} (t), x_{\rm N+1} = x_{\rm t}(t)$ depend on time. The continuum limit is obtained as
$N\rightarrow \infty$.} 
\label{discretized}
\end{figure}

\vskip 0.5cm
{\bf 6 Rubber block dynamics}

We consider the simplest but most fundamental case of a rubber block
(mass $M$) 
squeezed against a substrate, and with the upper surface clamped and in prescribed lateral
motion 
as indicated in Fig. \ref{discretized}. 

The shear stress that acts on the lower surface of the rubber block is treated 
as a spatially uniform stress $\sigma_{\rm f}(t)$.
In the most general case, where inertia effects are important,
we discretize the block (thickness $d$) 
into $N$ layers of thickness $\Delta d = d/N$ and mass $m=M/N$. 
The coordinates $x_1(t)= x_{\rm b}(t)$,
$x_2(t), ..., x_{\rm N}(t), x_{\rm N+1} (t) = x_{\rm t}(t)$ 
depend on time. The continuum limit is obtained as
$N\rightarrow \infty$.
The non-stationary sliding 
of the rubber block [see Fig. \ref{discretized}]
is determined by the equations of motion
$$m\ddot x_b(t) = F_1(t) -F_0[x_b(t)],$$
$$m\ddot x_2(t) = F_2(t)-F_1(t),$$
$$m\ddot x_3(t) = F_3(t)-F_2(t),$$
$$...$$
$$m\ddot x_{N}(t) = F_{N}(t)-F_{N-1}(t),\eqno(55)$$
where $F_0$ is the friction force acting on the bottom surface of the block,
and $F_i$ the shear force acting on layer $i$ from the layer $i+1$ above.
$F_{N}$ is the force acting on layer ${N}$ from the drive (the black slab
on top of rubber block in Fig. \ref{discretized}).

Note that the friction force $F_0[x_b(t)]$ is a {\it functional} of $x_b(t)$ since it
depends on the whole history $\{x_b(t'); t' \leq t\}$.
The shear force $F_i$ acting on layer $i$ from the layer above, can be 
derived from the equation relating the shear stress $\sigma (\omega)$ to the shear
strain $\epsilon (\omega)$ via 
$$\sigma(\omega ) = G(\omega ) \epsilon (\omega),\eqno(56)$$
where $G(\omega )$ is the complex frequency dependent shear modulus.
In our case ($i=1,...,N$) 
$$\epsilon (t) = [x_{i+1}(t)-x_{i}(t)]/\Delta d,\eqno(57)$$
where $\Delta d$ is the thickness of the rubber block.
If we multiply (56) with the area $A_0=L_xL_y$ of  
the tread block we get the shear force
$$F_i(\omega) = A_0 G(\omega) \epsilon (\omega),$$
or in time-space:
$$F_i(t) = \int d\omega \ F_i(\omega) e^{-i\omega t}$$ 
$$ = \int d\omega \ A_0 G(\omega ) \epsilon(\omega ) e^{-i\omega t}.\eqno(58)$$
Substituting 
$$\epsilon (\omega ) = {1\over 2 \pi} \int dt' \ \epsilon (t') e^{i\omega t'}$$
in Eq. (58) gives
$$F_i(t) = \int_{-\infty}^{\infty} dt' \ A_0 G(t-t') \epsilon(t'),\eqno(59)$$ 
where
$$G(t) = {1\over 2 \pi} \int_{-\infty}^{\infty} 
d\omega \ G(\omega) e^{-i\omega t}.\eqno(60)$$
If we assume that $\epsilon(t) = 0$ for $t<0$ and if we use that $G(t)$ must vanish
for $t<0$ because of causality (see Sec. 7), we get
$$F_i(t) = \int_0^t dt' A_0 G(t-t') \epsilon(t'),\eqno(61)$$
or, if we use (57),
$$F_i(t) = \int_0^t dt' \ K(t-t') [x_{i+1}(t')-x_{i}(t')],\eqno(62)$$
and in particular
$$F_{N}(t) = \int_0^t dt' \ K(t-t') [x_{\rm t}(t')-x_{N}(t')],\eqno(63)$$
where the {\it memory} spring
$$K(t)= A_0 G(t)/\Delta d.\eqno(64)$$

The equations above can also be applied when the properties of the rubber block depends on the
vertical coordinate $z$ if one replace the 
mass $m\rightarrow m_i$ and memory spring $K\rightarrow K_i$.
If $\rho =\rho(z)$ is the rubber mass density, then $m_i = A_0\Delta d \rho(z_i)$
and similarly  $K_i= A_0G(z_i,t)/\Delta d$.
Since the rubber friction $F_0$ depends extremely strongly on the sliding velocity
for very small sliding velocities, the system of differential equations given above is of the stiff
nature, which requires special care in the time-integration in order to avoid numerical instabilities.

\vskip 0.5cm
{\bf 7 Viscoelastic modulus}

Many experimental techniques can be used to obtain 
the complex, frequency dependent, shear modulus $G(\omega)$
[or Young modulus $E(\omega)$].
This quantity enters directly in the calculation of the rubber frictional
shear stress $\sigma_{\rm f}$. But in the 
rubber block dynamics theory described in Sec. 6 we also need the time-dependent {\it real} 
shear modulus
$$G(t) = {1\over 2 \pi} \int_{-\infty}^{\infty} 
d\omega \ G(\omega) e^{-i\omega t}.\eqno(65)$$
Since $G(\omega)$ usually is known only in a finite frequency interval, 
$\omega_0 < \omega < \omega_1$,
it is not easy to
calculate $G(t)$ directly from (65) using, e.g., the 
Fast Fourier Transform method. Another, even more serious problem 
is that in some cases $G(t)$ is not a ``perfect'' 
linear response function (see below), 
which gives rise to serious problems if one tries to calculate $G(t)$ directly from (65), e.g.,
one finds that $G(t)$ is complex rather than real.

\begin{figure}[htb]
\includegraphics[width=0.4\textwidth]{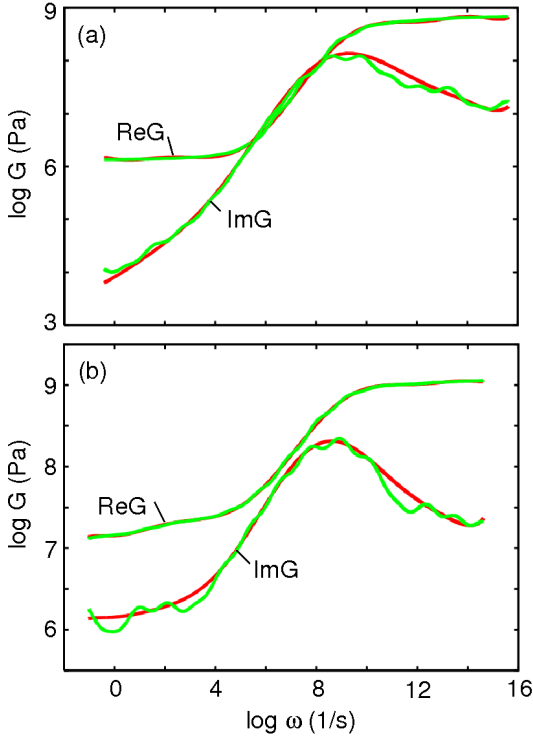}
\caption{
The real and the imaginary part of the viscoelastic modulus $G(\omega)$
for Styrene Butadiene copolymer (a) without filler, and (b) with $30 \%$ carbon black filler.
The red and green lines are the original data and the fit-data obtained by fitting 
the experimental $G(\omega)$ to a sum over relaxation times as described by (68). 
For the temperature $T=60 \ ^\circ C$.
} 
\label{FilledUnfilled}
\end{figure}

Let us first assume that $G(\omega)$ is the viscoelastic modulus
of a linear viscoelastic solid. 
Now, the shear stress $\sigma(t)$ in a solid at time $t$ can only depend on the
deformations (or strain) the solid has undergone at earlier times, i.e., it cannot 
depend on the future
strain ({\it causality}). 
Thus we can write                    
$$\sigma (t) = \int_{-\infty}^{t} dt' \ G(t-t') \epsilon (t').$$
The Fourier transform of this equation gives
$$\sigma (\omega) = G(\omega ) \epsilon (\omega)$$
where
$$G(\omega)= \int_0^\infty dt \ G(t) e^{i\omega t}$$
Since ${\rm Re} (i\omega t) < 0$ for  $t >0$ and ${\rm Im} (\omega) > 0$ it follows
that $G(\omega)$ is an analytical function of $\omega$ in the upper half of the complex frequency plane.
Thus all the singularities (poles and branch cuts) of $G(\omega)$ must occur in the lower part of
the complex $\omega$-plane and we may write 
$$G(\omega)=G_\infty-\int_0^\infty d\tau 
{H(\tau) \over 1-i\omega \tau}\eqno(66)$$
where the {\it spectral density} $H(\tau )$ is a {\it real} positive function
of the {\it relaxation time} $\tau$. This representation of $G(\omega)$ has the
correct analytical properties (all the singularities occur in the lower part
of the complex $\omega$-plane, as required by causality). The idea is now
to determine $G_\infty$ and $H(\omega)$ so that the difference $\Delta G(\omega)
=G_{\rm meas}(\omega)-G(\omega )$ between the measured $G_{\rm meas}$ shear
modulus and the analytical expression $G(\omega )$ given by (66) becomes as small
as possible. Thus, we minimize the quantity  
$$V= {1\over \omega_1-\omega_0} 
\int_{\omega_0}^{\omega_1} d\omega \ {|\Delta G(\omega)|\over |
G_{\rm meas}(\omega)|}.\eqno(67)$$
Note that
$$\Delta G = G_{\rm meas} (\omega) -              
G_\infty+\int_0^\infty d\tau {H(\tau) \over 1-i\omega \tau}$$
In practice, $G_{\rm meas}$ is only known at a set of discrete
frequencies $\omega_n$, $n=1,2,...,N$. Thus, (67) must be replaced with
$$V= {1\over N} \sum_{n=1}^N {|\Delta G(\omega_n)|\over |G_{\rm meas} (\omega_n)|}.$$
Furthermore, in numerical calculations it 
is only possible to include a finite number of 
relaxation times $\tau_k$ in the spectral representation of $G(\omega)$ and 
write
$$G(\omega)\approx G_\infty-\sum_k 
{H_k \over 1-i\omega \tau_k}.\eqno(68)$$
where
$$H(\tau) \approx \sum_k H_k \delta (\tau-\tau_k)$$
If the relaxation times are distributed exponentially (i.e., $\tau_k = \tau_0
{\rm exp}(\eta_k)$, where $\eta_k$ are uniformly distributed), it is usually enough
to include $\sim 10-15$ relaxation times, or one relaxation time for each decade in frequency.
If the measured $G_{\rm meas}(\omega)$, $\omega_0 < \omega < \omega_1$,
correspond to a true linear response function then one can find 
$G_\infty$ and $H(\omega)$ so that $V$ vanish. Since in many cases
$G(\omega)$ is not a true linear response function (see below) it is impossible 
to choose the real function $H(\omega)$ so that $V$ vanish. 
However, if we can find $H(\omega)$ and $G_\infty$ so that $V << 1$ we have a very good
representation of $G(\omega)$ of the form (66). 
Using (66) it is easy to integrate (65) to obtain 
$$G(t) = G_\infty \delta (t)-\theta (t)\int_0^\infty d\tau \ {H(\tau) \over \tau} e^{-t/\tau}$$
or, for a finite set of relaxation times,
$$G(t) = G_\infty \delta (t)-\theta (t)\sum_k {H_k \over \tau_k} e^{-t/\tau_k}$$
                               
The minimization of $V$ with respect to $G_\infty$ and $H_k$ 
can be performed by using a random number generator: 
First some (arbitrary) set of parameters $G_\infty$ and $H_k$ are chosen and $V$ calculated.
Next,
one replace $G_\infty$ and $H_k$
with $G_\infty (1 + \xi)$ and $H_k (1 +\xi_k)$, where $\xi$ and $\xi_k$ are small random numbers.
If the new $V$, calculated using these parameter values, is smaller than the original
one, then the new parameters are accepted as improved parameters; otherwise
they are rejected and the old parameters kept. 
If this procedure is repeated (iterated) many times the effective ``potential'' $V$ will converge
towards a minima, and the parameters $G_\infty$ and $H_k$ at the minima
of $V$ are the best possible ones, which we use in calculating $G(t)$ as described above. 

\begin{figure}[htb]
\includegraphics[width=0.4\textwidth]{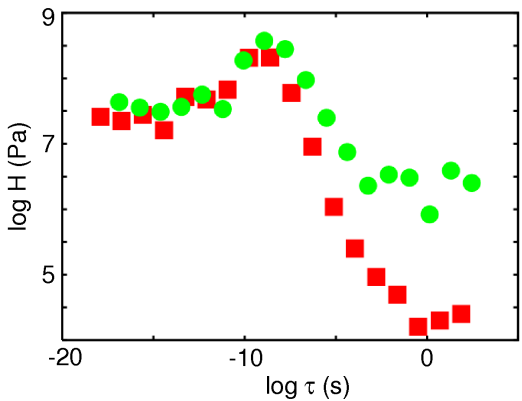}
\caption{
The spectral density $H_k$ as a function of the relaxation time $\tau_k$ as obtained
by fitting the measured $G(\omega)$ to the spectral representation (68). 
The squares and circles correspond to the unfilled and filled SB copolymer.
For the temperature $T=60 \ ^\circ C$.
} 
\label{tau.H}
\end{figure}

As an example, in Fig. \ref{FilledUnfilled}(a) we show 
the real and the imaginary part of the viscoelastic modulus $G(\omega)$
for Styrene Butadiene copolymer (a) without filler, and (b) with $30 \%$ carbon black filler.
The red and green lines are the experimental data and the fit-data, respectively,
where the fit-data was obtained by approximating
the experimental $G(\omega)$ by the sum (68) as described above. Note that
for the unfilled rubber
{\it both} the real and imaginary part of $G(\omega)$ are very well fitted by suitable choice of a
single {\it real} (positive) function $H(\tau)$. This is possible only because 
${\rm Re} G(\omega)$ and 
${\rm Im} G(\omega)$ are not independent but related via a Kramers-Kronig relation (see below).
For the filled rubber the fit of ${\rm Re} G$ is equally good, but there is a small
deviation for ${\rm Im} G$. We have observed the same for other rubber compounds, and sometimes
the deviation $\Delta G$ between measured and fitted ${\rm Im} G$ is 
larger than in Fig.  \ref{FilledUnfilled}(b),
but the calculated $G(t)$ is still accurate
enough for our applications. The deviation $\Delta G$ reflects
the fact that for filled rubber $G(\omega )$ is not 
a ``perfect'' linear response functions, but exhibit non-linearly.
One source of non-linearity is the 
so called Payne effect, due to the
strain-induced break-up of the filler network\cite{Payne,Gert}.

Fig. \ref{tau.H} shows the spectral density $H_k$ as a function of the 
relaxation time $\tau_k$ as obtained
by fitting the measured $G(\omega)$ to the spectral representation (68).
The squares and circles correspond to the unfilled and filled SB copolymer.
Note that the main difference between the two cases occur for 
the longest relaxation times $\tau > 10^{-8} \ s$, where the filled
rubber has a much higher spectral density $H(\tau)$. 
This implies that for the filled rubber, in contrast to
unfilled rubber, there
is a high spatial density of (relatively) high-energy-barrier 
rearrangement processes.
This results from polymer molecules bound to the surfaces
of the carbon filler particles; the bound layers, which may be $\sim 1 \ {\rm nm}$ thick,  
are in a glassy-like state and exhibit larger energy barriers for 
(thermally activated) rearrangements, 
as compared to the polymer segments further away from the filler particles.

We emphasize that only when
$G(\omega )$ corresponds to a linear response function causality will result
in a well-defined analytical structure for $G(\omega )$.
In this case the real part $G_1= {\rm Re} G$ 
and imaginary part $G_2= {\rm Im} G$
of $G(\omega)$ are related via Kramers-Kronig relations:
$$G_1(\omega)-G_\infty = {1\over \pi} P \int_{-\infty}^\infty d\omega' \ {G_2(\omega')\over
\omega'-\omega},\eqno(69)$$
$$G_2(\omega)= - {1\over \pi} P \int_{-\infty}^\infty d\omega' \ {G_1(\omega')-G_\infty \over
\omega'-\omega},\eqno(70)$$
where $G_\infty = G(\infty )$ is the (real) high frequency shear modulus.
For filled rubbers, (69) and (70) only hold approximately. In particular, due to the Payne effect 
(see above) the rubber behaves in a non-linear way
i.e., $G(\omega)$ is not a ``perfect'' linear response function, and
eqs. (69) and (70) will only hold approximately.

\begin{figure}[htb]
\includegraphics[width=0.4\textwidth]{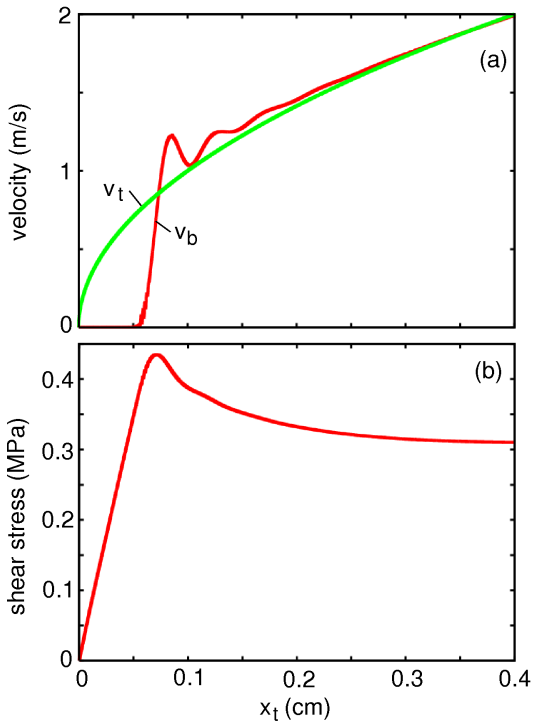}
\caption{
A rubber block with thickness $0.5 \ {\rm cm}$ sliding on
an asphalt road. The nominal pressure $\sigma_0 = 0.4 \ {\rm MPa}$.
The upper surface of the rubber block is clamped and undergoes
uniform acceleration $a= 500 \ {\rm m/s^2}$ for $0.004 \ {\rm s}$.
(a) The velocities $v_{\rm t}$ and $v_{\rm b}$ of the top and bottom
surface of the rubber block, respectively, as a function of the distance $x_{\rm t}$
the top surface has moved.
(b) The shear stress acting on the bottom surface of the rubber block,  as a function of $x_{\rm t}$.
In the calculation the block is ``divided'' into $N=12$ layers.} 
\label{xt.vel.stress}
\end{figure}

\begin{figure}[htb]
\includegraphics[width=0.4\textwidth]{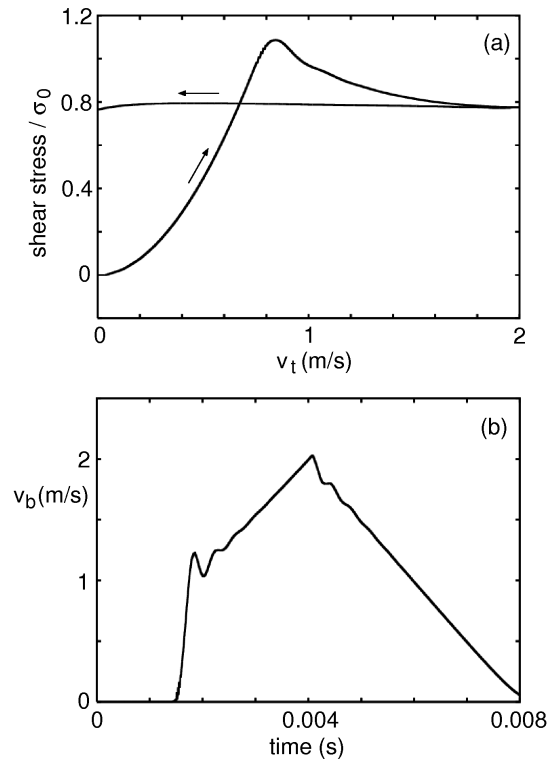}
\caption{
A rubber block with thickness $0.5 \ {\rm cm}$ sliding on
an asphalt road. The nominal pressure $\sigma_0 = 0.4 \ {\rm MPa}$.
The upper surface of the rubber block is clamped and undergoes
uniform acceleration $a= 500 \ {\rm m/s^2}$ for $t<0.004 \ {\rm s}$,
and then uniform retardation 
$a= -500 \ {\rm m/s^2}$ 
for $0.004 <t< 0.008 \ {\rm s}$.
(a) The shear stress, divided by the nominal pressure $\sigma_0$, 
acting on the bottom surface of the rubber block,  
as a function of velocity $v_{\rm t}$ of the top surface of the block.
(b) The velocity $v_{\rm b}$ of the bottom
surface of the rubber block, as a function of time.
In the calculation the block is ``divided'' into $N=12$ layers.} 
\label{up.down}
\end{figure}

\vskip 0.5cm
{\bf 8 Numerical results: rubber block dynamics}

In this section we present results for the sliding dynamics of
a rubber block with thickness $0.5 \ {\rm cm}$, squeezed against
an asphalt road with the nominal pressure $\sigma_0 = 0.4 \ {\rm MPa}$.

Assume first that 
the upper surface of the rubber block is clamped and undergoes
uniform acceleration $a= 500 \ {\rm m/s^2}$ for $0.004 \ {\rm s}$.
Fig. \ref{xt.vel.stress} shows 
(a) the velocities $v_{\rm t}$ and $v_{\rm b}$ of the top and bottom
surface of the rubber block, respectively, 
and (b) the shear stress acting on the bottom surface of the rubber block,  
as a function of the distance $x_{\rm t}$
the top surface has moved.
Note that the bottom surface of the block is effectively pinned until the shear stress
reaches $\sigma_{\rm s} \approx 0.44 \ {\rm MPa}$, corresponding to the friction coefficient
$\mu = \sigma_{\rm s} /\sigma_0 \approx 1.1$. After the depinning, the velocity of the bottom
surface of the block increases towards the velocity of the top surface, while the shear
stress approaches a constant value determined by the kinetic friction coefficient. The physical
reason for the peak in the shear stress at $x_{\rm t} \approx 0.07 \ {\rm cm}$ is due to
the flash temperature; the full flash temperature is not built up until the slip distance is of
order the diameter of the macro asperity contact regions, i.e., of order 0.4 cm in 
the present case.

Next, let us consider a case when the 
upper surface of the rubber block first
accelerates with $a= 500 \ {\rm m/s^2}$ for $0<t<0.004 \ {\rm s}$,
and then retards with 
$a= -500 \ {\rm m/s^2}$ 
for $0.004 <t <0.008 \ {\rm s}$.
In Fig. \ref{up.down}(a) we show
the shear stress, divided by the nominal pressure $\sigma_0$, 
acting on the bottom surface of the rubber block,  
as a function of velocity $v_{\rm t}$ of the top surface of the block.
In  Fig. \ref{up.down}(b)
we show the velocity $v_{\rm b}$ of the bottom
surface of the rubber block, as a function of time.

\begin{figure}[htb]
\includegraphics[width=0.4\textwidth]{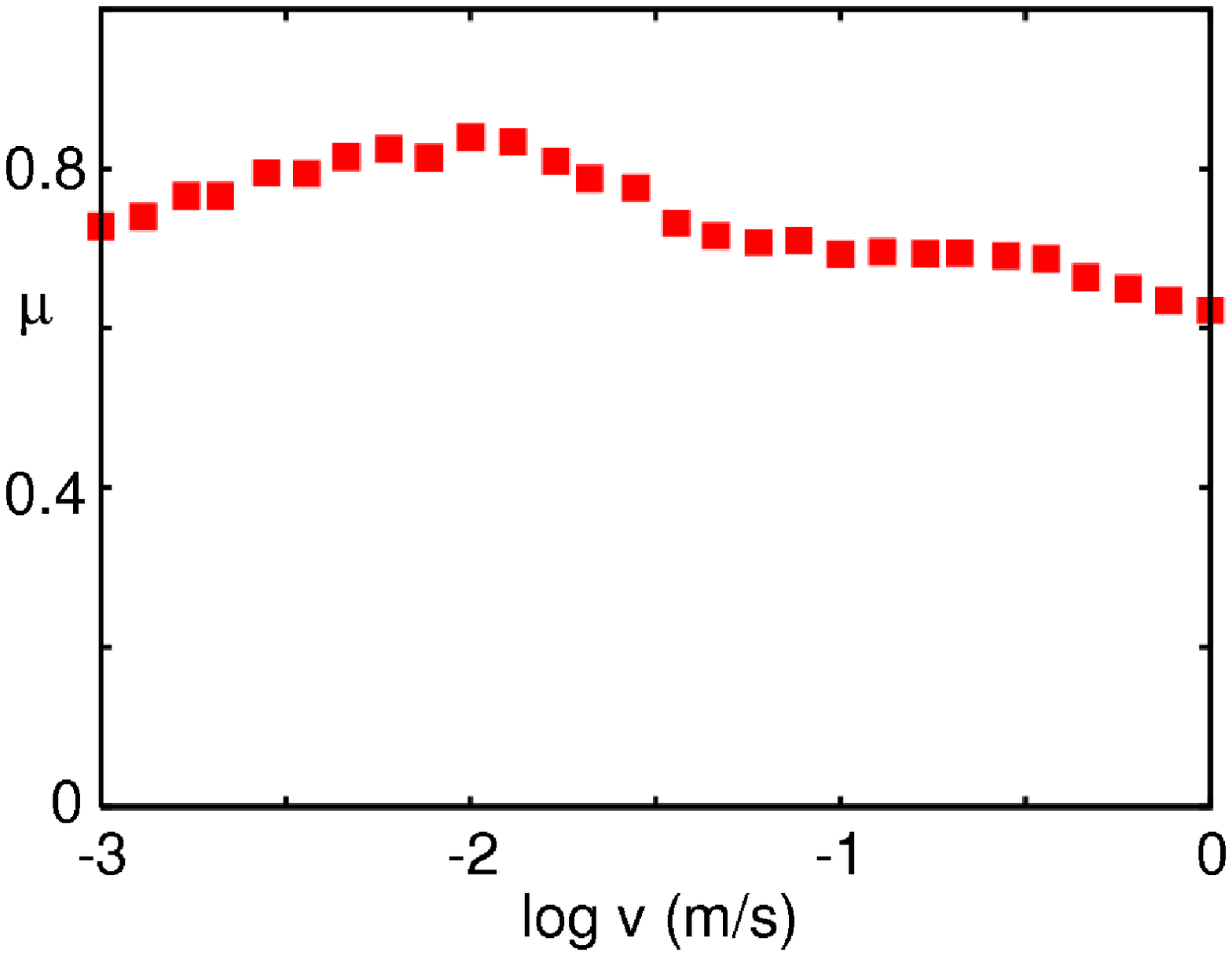}
\caption{
The friction coefficient $\mu$ as a function
of the logarithm of the sliding velocity.
For a rubber tread block sliding on an 
asphalt road surface at $T=18 \ ^\circ {\rm C}$.
(Courtesy of Olaf Lahayne).
} 
\label{exp99}
\end{figure}

\begin{figure}[htb]
\includegraphics[width=0.4\textwidth]{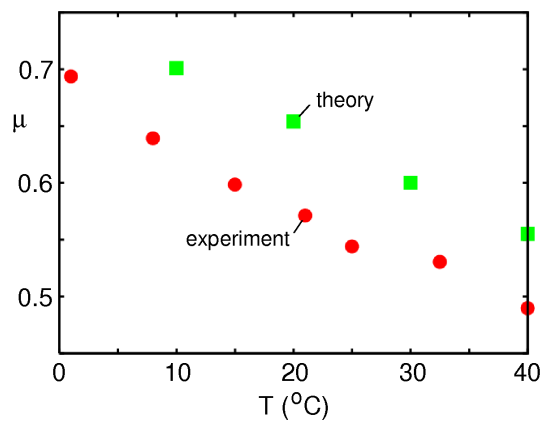}
\caption{
The friction coefficient $\mu$ as a function
of the background temperature, during sliding
of a rubber block on a rough substrate. The experimental data
was obtained using a portable skid tester\cite{test1,test2}
for tread rubber on road surface. The theory data was calculated
using the viscoelastic modulus of a standard tread rubber, sliding on
an asphalt road surface (with the power spectrum given 
in Fig. \ref{CqOpel}, surface 1)
at $v=1 \ {\rm m/s}$. 
} 
\label{exp2}
\end{figure}

\begin{figure}[htb]
\includegraphics[width=0.5\textwidth]{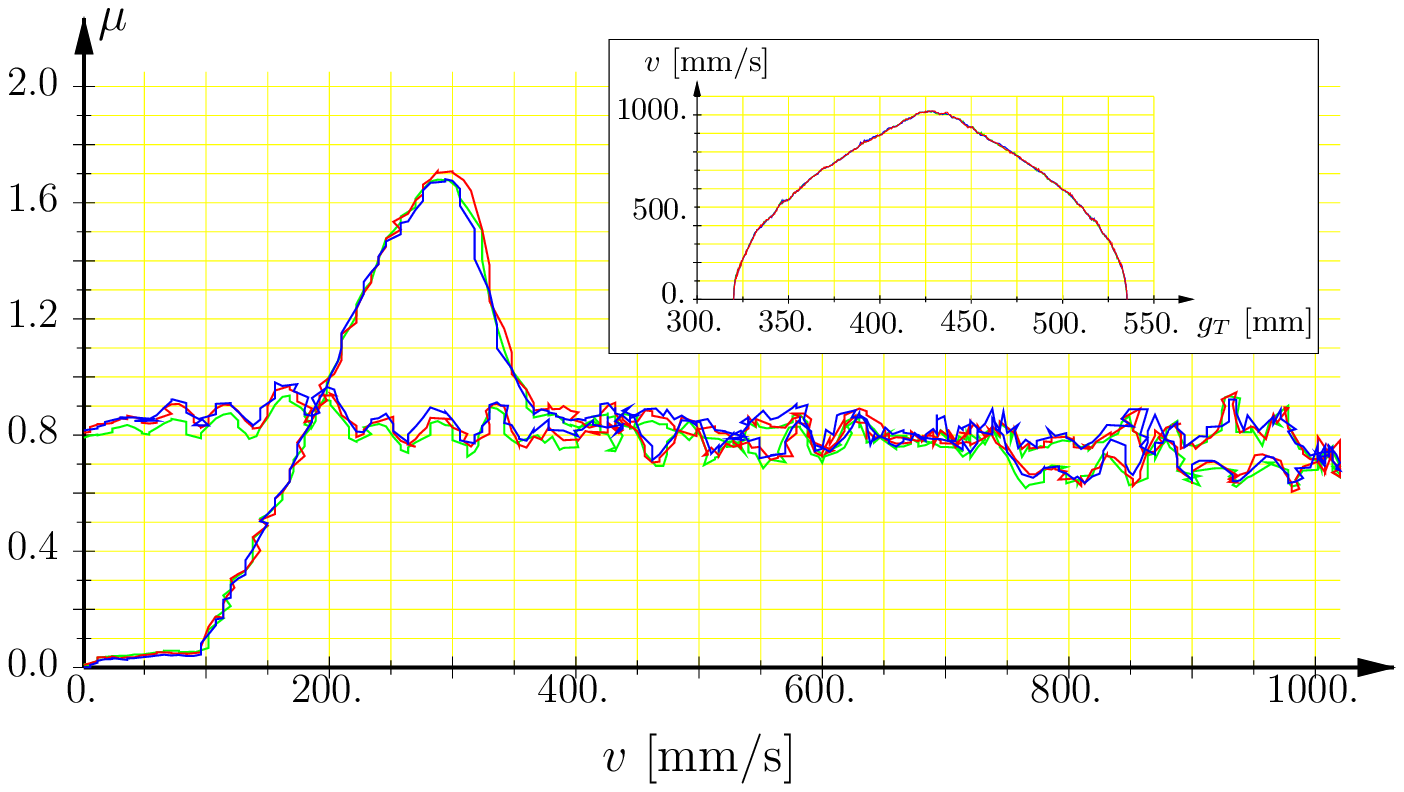}
\caption{
A soft rubber block (thickness $1.1 \ {\rm cm}$) sliding on a wet concrete surface at the nominal
pressure $0.2 \ {\rm MPa}$.
The variation of the effective friction coefficient 
with the velocity $v$ of the drive is shown for non-uniform sliding (see inset) involving
acceleration ($a \approx 5 \ {\rm m/s^2}$) followed by retardation
($a \approx -5 \ {\rm m/s^2}$). From Ref. \cite{KGK} (with permission).
} 
\label{exp1}
\end{figure}

\vskip 0.5cm
{\bf 9 Comparison with experiment}

A lot of experimental data has been presented in the literature related to rubber friction.
However, in order to quantitatively compare the rubber friction theory with 
experimental data, both the viscoelastic modulus $E(\omega,T)$ and the substrate
surface roughness power spectrum $C(q)$ must be known, but 
this information was never reported in any experimental
investigations of rubber friction which I am aware of. Thus, in the present section we will
only consider general (universal) aspects of rubber friction, and show that the theory is in
good qualitative agreement with the presented experimental data.

Let us first consider stationary sliding. Fig. \ref{exp99} shows the 
measured kinetic rubber friction coefficient
as a function of the logarithm of the sliding velocity,
for a tread rubber block sliding on an asphalt surface\cite{Olaf}.
The velocity dependence of $\mu_{\rm k}(v)$ is in good 
qualitative agreement with the theory (compare Fig. \ref{exp99} with Fig. \ref{mtav}(a)). 
More generally,  
I have found that the theory, and all the experiments known to me, 
gives a maximal friction coefficient of order unity,
and the position of the maximum in the range $10^{-3}-10^{-1} \ {\rm m/s}$,
depending on the rubber compound and the substrate surface. In the absence of the flash temperature,
according to the theory 
the maximum would instead occur at much higher sliding velocities, typically in the range
$10^2-10^4 \ {\rm m/s}$. This illustrates the crucial role of the flash temperature.

Next, let us consider the dependence of the friction 
coefficient $\mu$ 
on the background temperature.
Fig. \ref{exp2} shows experimental data
(obtained using a portable skid tester\cite{test1,test2})
for a tread rubber block sliding on a road surface. 
The theory data in the same figure is calculated for a rubber block
sliding on
an asphalt road surface (with the power spectrum given 
in Fig. \ref{CqOpel}, surface 1)
at $v=1 \ {\rm m/s}$. 
The 
viscoelastic modulus of the rubber used in the calculation is for a tread rubber. 
As expected, the rubber friction decreases with increasing
background temperature. 

\begin{figure}[htb]
\includegraphics[width=0.4\textwidth]{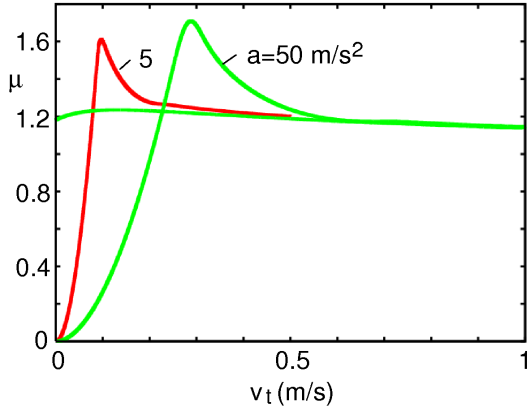}
\caption{
A rubber block (thickness $0.5 {\rm cm}$) sliding on an asphalt road surface
at the pressure $0.2 \ {\rm MPa}$.
The (calculated) variation of the effective friction coefficient 
with the velocity $v_{\rm t}$ is shown for two different cases where
the drive first  
accelerate ($a \approx 5 \ {\rm m/s^2}$ and $50 \ {\rm m/s^2}$) followed by retardation
($a \approx -5 \ {\rm m/s^2}$ and $-50 \ {\rm m/s^2}$).
} 
\label{calc1}
\end{figure}

Let us now consider non-stationary sliding.
Fig. \ref{exp1} shows 
results for a (soft) rubber block (thickness $1.1 \ {\rm cm}$) sliding on a 
wet concrete surface at the nominal
pressure $0.2 \ {\rm MPa}$.
The variation of the effective friction coefficient 
with the velocity $v$ of the drive is shown for non-uniform sliding (see inset) involving
acceleration ($a \approx 5 \ {\rm m/s^2}$) followed by retardation
($a \approx -5 \ {\rm m/s^2}$)\cite{KGK}.
Note that the experimental data is of the general form predicted by the theory 
(see Fig. \ref{calc1}) with a ``start-up'' peak due to the flash-temperature; the rubber block must slide a
distance of order the linear size of the macro asperity contact regions before the full flash temperature
has been developed, and this is the origin of the ``start-up'' peak.

Note that the ``start-up'' peak in Fig. \ref{exp1} and in the calculations \ref{calc1} has nothing to do
with the static friction coefficient, but rather is a kinetic effect related to the
finite sliding distance necessary in order to fully build up the
flash temperature. In fact, rubber on rough substrates does not
exhibit any static friction coefficient, but only a kinetic friction which 
(for small sliding velocities) decreases continuously with decreasing sliding velocity. 

\begin{figure}[htb]
\includegraphics[width=0.4\textwidth]{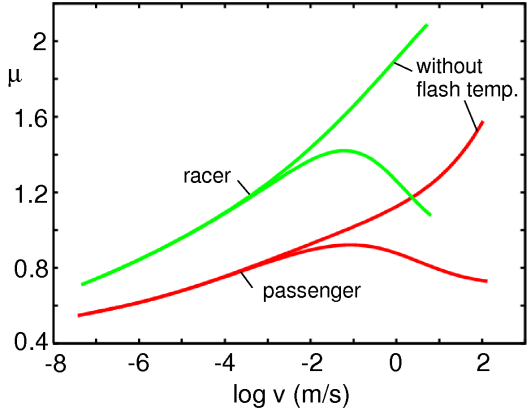}
\caption{
The kinetic (steady state) rubber friction coefficient for typical tire tread
compounds for racer car (top) and personal car (bottom), sliding on
an asphalt road surface.
The background temperature $T=60 \ {\rm ^\circ C}$.
} 
\label{exp3}
\end{figure}

Finally, let us check if the theory agrees with the observation that tires for racer cars
(or racer motorcycles) exhibit much larger friction (but also much larger wear) than tires for
passenger cars. In Fig. \ref{exp3} I show the calculated kinetic friction 
coefficient, using the measured viscoelastic
modulus of a
rubber compound from racer tires, and for a passenger car tire, assuming all other conditions identical. 
The maximum kinetic friction is about $50 \%$ higher for the racer compound which is also what
is typically found experimentally\cite{RacerPaul}. This calculation indicates that even for racer tires
the main origin of the friction is due to the internal damping of the rubber rather than
an adhesive tire-road interaction.

\vskip 0.5cm
{\bf 10 On the origin of the short-distance cut-off}

The rubber friction theory presented above assumes that the friction is due
to the internal friction of the rubber. That is, the substrate surface asperities
exert forces on the rubber surface and result in pulsating
deformations of the rubber. The deformations cannot occur completely adiabatically,
but result in transfer of  
energy into random heat motion in the 
rubber. In calculating this asperity-induced contribution to the friction I include
only the road surface roughness with wavevectors $q<q_1$. 
Here I will discuss the origin of the {\it short
distance} cut off $q_1$.
In principle, there may be several different origins of 
$q_1$. For example, if the road is covered by small particles, e.g., dust or
sand particles, with typical diameter $D$, then one may expect $q_1 \approx 1/D$.
Similarly, on a wet road, the water trapped in the surface cavities 
may act as an effective short-distance cut off\cite{NatureM}. 
However, for clean dry road surfaces, 
I believe that
the cut off may be determined by the rubber 
compound properties.

If no short-distance cut off $q_1$ would exist the flash temperature and the
surface stresses in the contact areas would increase as we study
the contact regions at higher and higher magnification. However, when the
flash temperature and surface stresses becomes high enough, the rubber will
degrade\cite{Ueba}. In fact, measurements on bulk samples have shown that natural rubber
rapidly thermally degrades already
at $T \approx 200 \ {\rm C}$. The surface layer of the tire rubber is exposed to
oxygen and ozone and will  
degrade even faster than in the bulk, even if the
temperature would be the same. The thermal degradation results in a thin layer of modified rubber
at the surface, and we will make the basic assumption that the deformation of the rubber
on length scales shorter than the thickness of the modified layer, gives a negligible
contribution to the tire-road friction. Thus, with respect to friction, we assume that the modified
surface layer acts as a ``dead'' layer.  

The formation of a modified surface layer
is a thermally activated mechanic-chemical process
involving the breaking of
chemical bonds (and the formation of new bonds). The rate of bond-breaking 
(at the temperature $T$ and the tensile stress $\sigma$) is
assumed to be given by the standard expression of activated processes:  
$$w \approx w_0 e^{-\Delta E (1- \sigma/\sigma_c )/k_BT}\eqno(71)$$
where the attempt frequency
$w_0$ is usually of order $10^{12} \ {\rm s}^{-1}$, and
where $\Delta E$ is the activation energy for bond breaking, and
$\sigma_c$ the stress necessary for bond-breaking at zero temperature. 
The weakest (chemical) bonds in rubber cross-linked with sulfur, are multi-sulfur
cross-links for which $\Delta E$ is of order $\sim 2-3 \ {\rm eV}$. 
Assuming a typical cross-link density, and that the applied stress 
distributes itself entirely on the cross-links 
(which requires so high temperature that the rubber is 
liquid-like between the cross-links), and assuming
that the same force acts on all the crosslinks, one can estimate
the stress $\sigma_c$ to be of order $\sim 1 \ {\rm GPa}$. 
However, in inhomogeneous materials such as rubber, there will be a large distribution of
local forces acting on the cross-links, and the cross-links where the local stress is highest 
will in general break first.

The contact time
of a tire tread block in the footprint area is typically of order $0.01 \ {\rm s}$,
and since the dead layer has been worn out in about $\sim 100$ contacts, 
the rate $w$ must be of order $1 \ {\rm s}^{-1}$. Taking
$w\sim 1 \ {\rm s}^{-1}$ gives
$$\Delta E (1-\sigma/\sigma_c )/k_BT = {\rm ln}(w_0/w) \approx 28\eqno(72)$$
Note that the RHS of this expression is not very sensitive to the value of $w$.
We can use (72) as a criteria to determining the optimum cut-off $q_1$:
We determine $q_1$ so that Eq. (72) is satisfied, where $T$ is now the flash
temperature in the contact area and $\sigma$ the stress in the contact area.

We have found that using $\Delta E \approx 1.25 \ {\rm eV}$ 
and $\sigma_c \approx 100 \ {\rm MPa}$ result in a friction coefficient in
good agreement with experiment. 
Note that
$\Delta E$ is smaller than the typical energy to break sulfur cross link (which is of order
$2-3 \ {\rm eV}$) or the energy to break a C-C bond along a carbon chain (which is of order
$3-4 \ {\rm eV}$). However, 
in the present case the bond breaking is likely to involve reaction with foreign molecules,
e.g., ozone or oxygen, 
and the effective activation energy for such processes 
may be much smaller than when the bond break
in vacuum. The value for $\sigma_c$ is higher than the rupture stress of macroscopic rubber blocks,
but $\sigma_c$ refer to the rupture stress of very small rubber volume elements,
which is higher than for macroscopic rubber blocks due to the absence of 
(large) crack-like defects\cite{explain}. 
It is clear that the processes which determines the cut-off $q_1$
are closely related to tire tread wear.

\vskip 0.5cm
{\bf 11 Summary and conclusion}

When a rubber block slips on a hard rough substrate, the substrate
asperities will exert time-dependent deformations of the rubber surface
resulting in viscoelastic energy dissipation in the rubber, which gives a contribution
to the sliding friction. Most surfaces of solids have roughness on many different length scales,
and when calculating the friction force it is necessary to include the viscoelastic deformations
on all length scales. The energy dissipation will result in local heating of the rubber.
Since the viscoelastic properties of rubber-like materials 
are extremely strongly temperature dependent, it is
necessary to include the local temperature increase in the analysis. 
In this paper I have developed a theory which describe the influence of the {\it flash temperature} 
on rubber friction.
At very low sliding velocity
the temperature increase is negligible because of heat diffusion, but already for velocities of order
$10^{-2} \ {\rm m/s}$ the local heating may be very important, and 
I have shown that in a typical case the temperature increase
result in a decrease in rubber friction  
with increasing sliding velocity for $v > 0.01 \ {\rm m/s}$. This may result in stick-slip
instabilities, and is of crucial importance in many practical applications, e.g., for the tire-road
friction, and in particular for ABS-breaking systems.

\vskip 0.3cm
{\bf Acknowledgement}
I thank O. Albohr for many useful discussion. I also thank Pirelli Pneumatici for support.

\vskip 1cm

\begin{figure}[htb]
\includegraphics[width=0.35\textwidth]{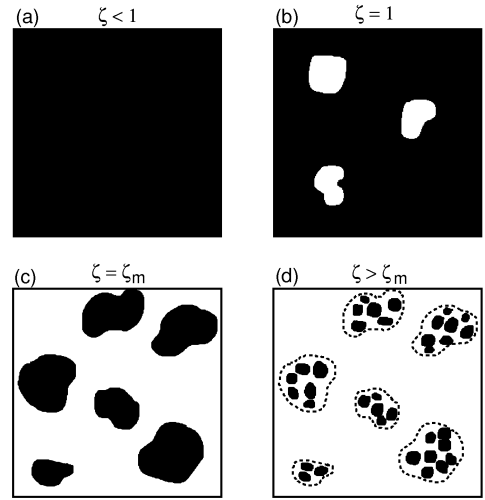}
\caption{
The contact area at increasing magnification (a)-(d). The macro-asperity
contact area (c) breaks up into smaller contact areas (d) 
as the magnification is increased.
} 
\label{newnew1}
\end{figure}

{\bf Appendix A: Average size of a macro-asperity contact region}

Consider the contact between two solids with nominally flat surfaces.
Fig. \ref{newnew1} shows
the contact area at increasing magnification (a)-(d). 
At low magnification $\zeta < 1$ it appears as if complete contact occur between the solids.
The macro-asperity contact regions typically appear for a magnification 
somewhere in the range $\zeta_{\rm m} =q_{\rm m}/q_0 \approx
2-5$ depending on the substrate surface and the rubber compound,
where $A/A_0 \approx 0.25-0.3$ is slightly below the site percolation threshold
(for a hexagonal lattice).
The macro-asperity
contact regions (c) breaks up into smaller contact regions (d) 
as the magnification is increased.

Let us now estimate the (average) size of the macro-asperity contact regions.
Assume that the contact patches have the surface areas $A_i$ ($i=1,...,N$).
Let us define the probability distribution
$$P_A = {1\over N} \sum_i \delta (A-A_i)$$
where $N$ is the number of contacting asperities.
The average area of a contact patch
$$ a = \int dA \ P_A A =
{1\over N} \sum_i A_i\eqno(A1)$$
Assume that the distribution $P_h$ of summit heights is a Gaussian,
$$P_h = {1\over (2\pi )^{1/2} h^*} {\rm exp} \left [-h^2/2{h^*}^2\right ],\eqno(A2)$$
where $h^*$ is the root-mean-square amplitude of the summit height fluctuations.
Assume that we can neglect the (elastic) interaction between the macro-contact areas.
Thus, using the Hertz contact theory 
the (normalized) area of real contact is \cite{[4],[5]}
$${\Delta A\over A_0} = \pi n_0 R_0 \int_d^\infty dh \ (h-d)P_h\eqno(A3)$$
where $A_0$ is the nominal contact area, $R_0$ the radius of curvature of the asperity,
$n_0$ 
the number of asperities per unit area, and $d$ the separation between the surfaces. 
The number of contacting asperities per unit area
$${N\over A_0} = n_0 
\int_d^\infty dh \ P_h\eqno(A4)$$
Substituting (A2) in (A3) and (A4), and introducing 
the new integration variable $\xi = (h-d)/h^*$ 
gives
$${\Delta A\over A_0} =  
n_0 R_0h^* 
\left ({\pi\over 2}\right )^{1/2} 
\int_0^\infty d\xi \ \xi 
{\rm exp} \left [-{1\over 2}\left 
(x+\xi\right )^2\right ]\eqno(A5)$$
and
$${N \over A_0} =  
{n_0 \over (2\pi)^{1/2} } 
\int_0^\infty d\xi \  
{\rm exp} \left [-{1\over 2}\left (x+\xi\right )^2\right ]\eqno(A6)$$
where $x=d/h^*$.
Thus, the average macro-asperity contact area
$$a = {\Delta A \over N} = {\pi R h^* 
\int_0^\infty d\xi \ \xi 
{\rm exp} \left [-{1\over 2}\left (x+\xi\right )^2\right ] \over
\int_0^\infty d\xi \
{\rm exp} \left [-{1\over 2}\left (x+\xi\right )^2\right ]}\eqno(A7)$$ 

We can estimate the concentration of asperities, $n_0$, the (average) radius of curvature
$R$ of the asperities, and the rms summit height fluctuation $h^*$ as follows.
We expect the asperities to form a hexagonal-like (but somewhat disordered)
distribution with the lattice constant $\lambda_{\rm m} = 2 \pi /q_{\rm m}$ so that
$$n_0 \approx 2/(\lambda_{\rm m}^2\surd 3) = {q_{\rm m}^2 \over 2 \pi^2 \surd 3} \approx 0.029 q_{\rm m}^2$$
The height profile along some axis $x$ in the surface plane will oscillate with the wavelength
of order $\approx \lambda_{\rm m}$ roughly as 
$h(x) \approx h_1 {\rm cos}(q_{\rm m} x)$ where $h_1 = \surd 2 h_0$ (where $h_0$ is the rms roughness amplitude).
Thus, 
$${1\over R} \approx h''(0) = q_{\rm m}^2 h_1 =\surd 2 q_{\rm m}^2 h_0.$$
Finally, we expect the rms of the fluctuation in the summit height to be somewhat
smaller than $h_0$.  
In what follows we use\cite{Nayak} 
$n_0 = 0.023q_{\rm m}^2$,
$1/ R_0 =0.92 h_0q_{\rm m}^2$, and
$h^* = 0.63 \ h_0.$ 
Using these results and defining $a=\pi R^2$ we obtain
from (A5) and (A7):
$${\Delta A\over A_0} =  
0.0198
\int_0^\infty d\xi \ \xi 
{\rm exp} \left [-{1\over 2}\left (x+\xi\right )^2\right ]\eqno(A8)$$

$$ (Rq_{\rm m})^2 = 0.688  { 
\int_0^\infty d\xi \ \xi 
{\rm exp} \left [-{1\over 2}\left (x+\xi\right )^2\right ] \over
\int_0^\infty d\xi \ 
{\rm exp} \left [-{1\over 2}\left (x+\xi\right )^2\right ]}\eqno(A9)$$ 

\begin{figure}[htb]
\includegraphics[width=0.45\textwidth]{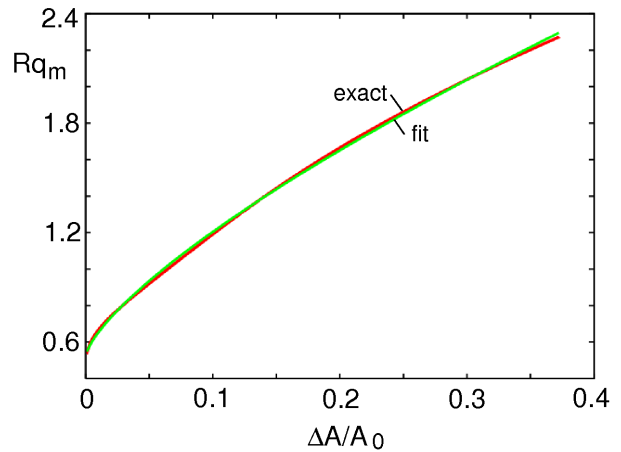}
\caption{
The (average) radius $R$ of a macro-asperity contact area as a function of the
(normalized) projected contact area $\Delta A/A_0$. 
} 
\label{newnew}
\end{figure}

In Fig. \ref{newnew} we show the radius $R$ of a macro-asperity contact region
as a function of the relative contact area $\Delta A/ A_0$.
The numerical data are well approximated by 
$$Rq_{\rm m} = a+b(\Delta A/A_0)^c\eqno(A10)$$
where $a=0.526$, $b=3.636$ and $c=0.729$.
This fit-function is also shown in Fig. \ref{newnew}. In a typical case when a tread
block is slipping on a road surface, $A(\zeta_{\rm m})/A_0=\Delta A/A_0 \approx 0.25$. 
Using Fig. \ref{newnew} this
gives 
the diameter of the macro-asperity contact regions $2R\approx 3/q_0$. If $q_0 = 600 \ {\rm m}^{-1}$
this gives $2R\approx 0.5 \ {\rm cm}$.

\end{document}